\documentclass[12pt,pre,aps,showpacs,preprint]{revtex4}

\usepackage{graphics}
\usepackage{graphicx}
\usepackage{amsfonts}
\usepackage{textcomp}
\usepackage{amssymb}
\usepackage{amsmath}

\begin{document}

\title{Optimal Packings of Superballs}

\author{Y. Jiao$^1$, F. H. Stillinger$^2$ and S. Torquato$^{2,3,4,5}$}


\affiliation{$^1$Department of Mechanical and Aerospace
Engineering, Princeton University, Princeton New Jersey 08544,
USA}



\affiliation{$^2$Department of Chemistry, Princeton University,
Princeton New Jersey 08544, USA}





\affiliation{$^3$Program in Applied and Computational Mathematics,
Princeton University, Princeton New Jersey 08544, USA}

\affiliation{$^4$School of Natural Sciences, Institute for
Advanced Study, Princeton NJ 08540}

\affiliation{$^5$Princeton Center for Theoretical Science,
Princeton University, Princeton New Jersey 08544, USA}

\date{\today}

\begin{abstract}

Dense hard-particle packings are intimately related to the
structure of low-temperature phases of matter and are useful
models of heterogeneous materials and granular media. Most studies
of the densest packings in three dimensions have considered
spherical shapes, and it is only more recently that nonspherical
shapes (e.g., ellipsoids) have been investigated. Superballs (whose shapes are defined
by $|x_1|^{2p}+|x_2|^{2p}+|x_3|^{2p}\le1$) provide a versatile family of convex particles ($p\ge
0.5$) with both cubic- and octahedral-like  shapes  as well as concave
particles ($0<p<0.5$) with octahedral-like shapes. In this paper, we
provide analytical constructions for the densest known superball packings
for all convex and concave cases. The candidate maximally dense packings
are certain families of Bravais lattice packings (in which each
particle has 12 contacting neighbors) possessing
the global symmetries that are consistent with 
certain rotational symmetries of a superball. We also provide strong evidence 
that our packings for convex superballs ($p \ge 0.5$) are most likely 
the optimal ones. The maximal packing density
as a function of $p$ is nonanalytic at the sphere-point ($p=1$)
and increases dramatically as $p$ moves away from unity.  Two more  
nontrivial nonanalytic behaviors occur at $p^*_c = 1.1509\ldots$ and 
$p^*_o=\ln3/\ln4=0.7924\ldots$ for ``cubic'' and ``octahedral'' 
superballs, respectively, where
different Bravais lattice packings possess the same densities. The
packing characteristics determined by the broken rotational
symmetry of superballs are similar to but richer than their
two-dimensional ``superdisk" counterparts [Y. Jiao, F. H.
Stillinger and S. Torquato, Phys. Rev. Lett., {\bf{100}}, 245504
(2008)], and are distinctly different from that of ellipsoid  packings.
Our candidate optimal superball packings provide a starting point
to quantify the equilibrium phase behavior of superball
systems, which should deepen our understanding of the statistical
thermodynamics of nonspherical-particle systems.

\end{abstract}


\maketitle

\section{Introduction}

Packing problems, such as how densely given solid objects can fill
$d$-dimensional Euclidean space $\Re^d$, have been a source of
fascination to mathematicians and scientists for centuries, and
continue to intrigue them today. Dense packings of hard particles
have served as useful models to understand the structures of
low-temperature phases of matter, such as liquids, glasses and
crystals \cite{Zallen,Chaikin,Torquato}, heterogeneous
materials \cite{Ki91,Torquato,Zo08}, granular media \cite{Edwards,Materials},
and even protein folding \cite{Di01}. Packing problems
arise in many different branches of pure mathematics \cite{Cassel, Conway}
and in information theory \cite{Sh48}.

In general, a collection of given solid objects (particles) in $d$-dimensional
Euclidean space $\mathbb{R}^d$ is called a packing if no two of
the objects have an interior point in common. The packing density
$\phi$ is defined as the fraction of space $\mathbb{R}^d$ covered by the particles.
A problem of great interest is the determination of the densest
arrangement(s) of such particles and the associated maximal
density $\phi_{max}$. Besides the aforementioned applications,
finding maximal density packings  is  of importance to 
understanding the structure and 
properties of crystalline equilibrium phases of particle systems 
as well as their ground-state ($T=0$) structures in low dimensions
in which the interactions are characterized by steep
repulsions and  short-ranged attractions.

The optimal solution to this packing problem for congruent
particles that do not tile space is already very challenging, even
in low-dimensional Euclidean spaces $\mathbb{R}^d$ ($d=2, 3$). In
two dimensions, the optimal packing of circular disks is the
well-known triangular-lattice packing with density $\phi_{max}=
\pi/(2\sqrt 3)$. For ellipses, the densest packing can be obtained
by an affine transformation of the triangular-lattice packing of
circular disks, which possesses the same density $\phi_{max}=
\pi/(2\sqrt 3)$. These packings can also be constructed by
enclosing each particle with a hexagon with minimum area that
tessellates $\mathbb{R}^2$ \cite{Fe64,Pach}. It is only recently that
the famous Kepler conjecture, which postulates that the densest
packing of spheres in $\mathbb{R}^3$ has a density
$\phi_{max}=\pi/(3\sqrt 2)$, as realized by the stacking variants
of the face-centered cubic (FCC) lattice packing, has been proved
\cite{Hales}. Although dense sphere packings  for $d \ge 4$ have
received considerable recent attention \cite{Cohn_Elkies, Parisi, SalExpMath, ParisiII, Scard},
optimal solutions are not rigorously known.
Packing problems in high Euclidean dimensions are intimately
related to the best way of transmitting digital signals over a
noisy channel \cite{Sh48,Conway}.

Understanding the organizing principles that lead to the densest
packings of nonspherical particles that do not tile space
is of great practical and fundamental interest. Clearly, the effect of asphericity
 is an important feature to include on the way to characterizing
more fully real dense granular media. Another important application
relates to supramolecular chemistry \cite{supra} of organic compounds
whose molecular constituents can possess many different types
of group symmetries \cite{Ki73}.
 
On the theoretical side, no results exist that rigorously prove the densest packings of other
congruent non-space-tiling particles in three dimensions. For
congruent ellipsoids, the densest known packings
with density $\phi_{max}= 0.7707\ldots$ are achieved by a family of
crystal packings (for a wide range of aspect ratios) in which each ellipsoid has contact with 14
others \cite{Alexks}. Recently, Conway and Torquato found dense periodic
packings of regular tetrahedra with $\phi \approx 0.72$ by filling
imaginary icosahedra with the
densest arrangement of 20 tetrahedra and arranging
the icosahedra  in their optimal Bravais lattice arrangement \cite{ConwaySal}. Chaikin et
al. experimentally produced jammed disordered packings of nearly
tetrahedral dice with density $\phi \approx 0.75$ \cite{Paul}.
Chen has discovered the densest known periodic packing of
tetrahedra with $\phi=0.7786\ldots$ \cite{BChen,prelim}.

\begin{figure}
\begin{center}
$\begin{array}{c@{\hspace{0.75cm}}c@{\hspace{0.75cm}}c@{\hspace{0.75cm}}c}
\includegraphics[height=3.0cm, keepaspectratio]{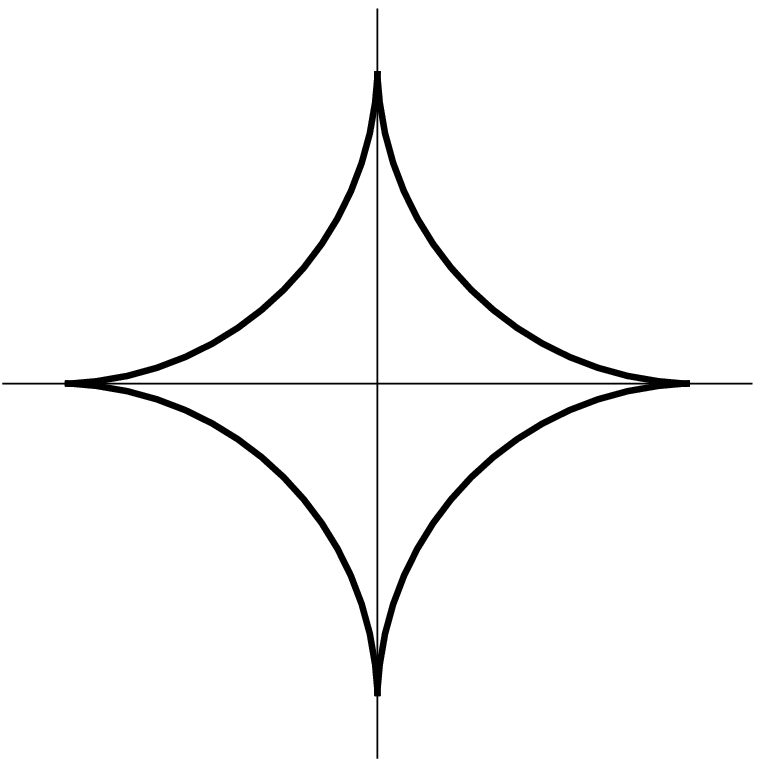} &
\includegraphics[height=3.0cm, keepaspectratio]{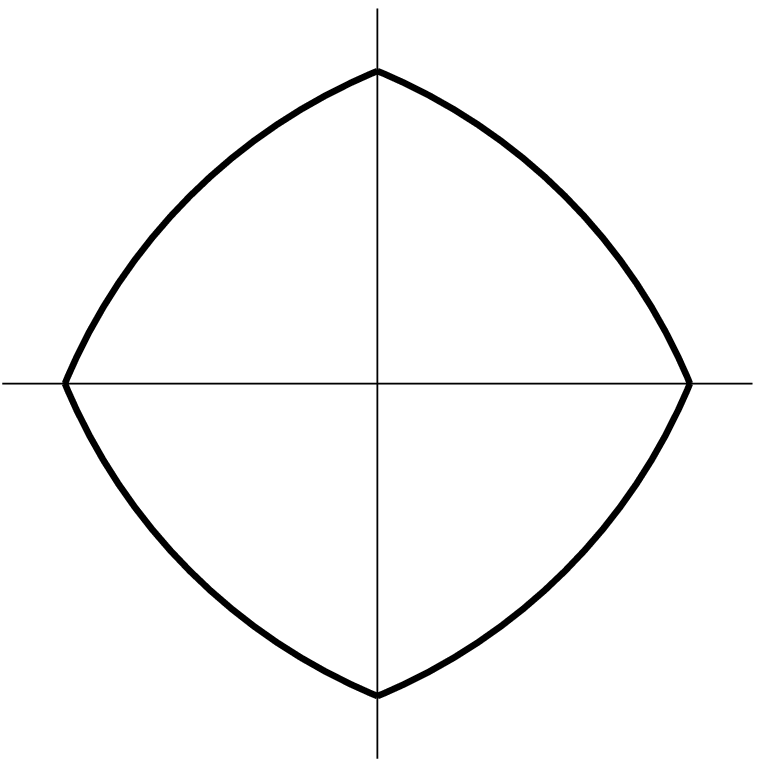} &
\includegraphics[height=3.0cm, keepaspectratio]{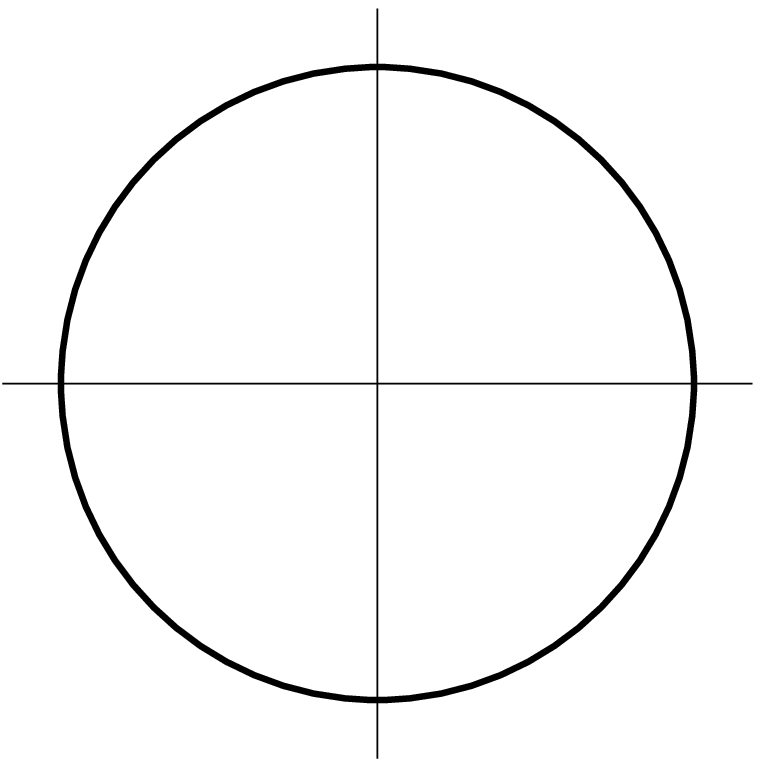} &
\includegraphics[height=3.0cm, keepaspectratio]{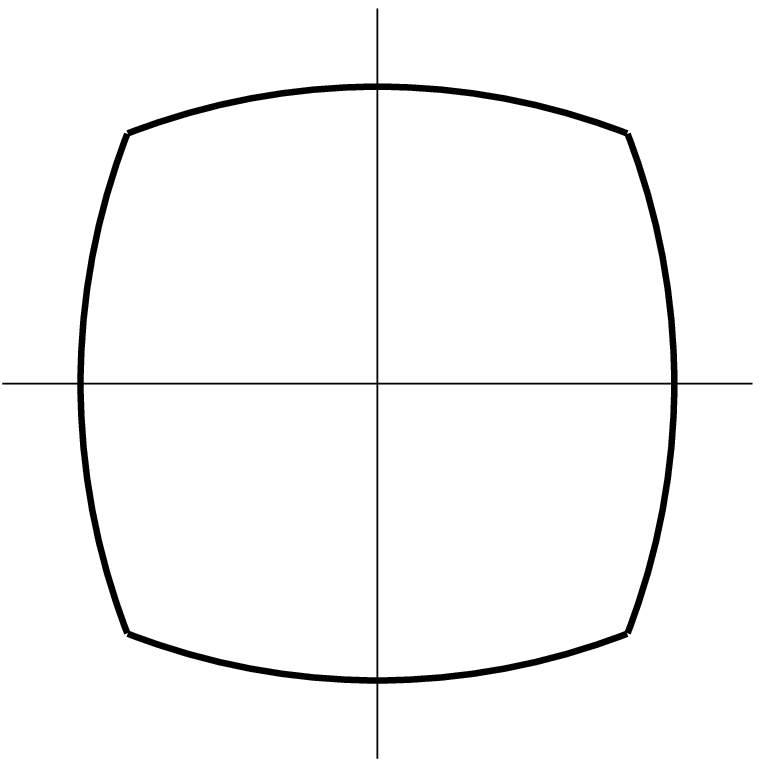} \\
\mbox{(a) p = 0.45} & \mbox{(b) p = 0.75} & \mbox{(c) p = 1.0} &
\mbox{(d) p = 2.0}
\end{array}$
\end{center}
\caption{Superdisks with different values of the deformation parameter $p$.}
\label{fig1}
\end{figure}

This paper is concerned with superball packings in low dimensions,
although primarily in three dimensions.
 A $d$-dimensional \textit{superball} is a centrally symmetric body in $\mathbb{R}^d$
occupying the region  \cite{Elkies}
\begin{equation}
|x_1|^{2p}+|x_2|^{2p}+\cdots+|x_d|^{2p} \le 1,
\end{equation}
where $x_i$ $(i=1,\ldots,d)$ are Cartesian coordinates
and $p \ge 0$ is the \textit{deformation parameter}, which
indicates to what extent the particle shape has deformed from that
of a $d$-dimensional sphere ($p=1$).
A particle is centrally symmetric if it has a center $P$ that
bisects every chord through $P$ connecting any two boundary points
of the particle. We note that a  particle is convex if the entire line segment
connecting two points of the particle also belongs to the
particle; otherwise it is concave. 
The terms \textit{superdisk} and \textit{superball} will be our designations
for the two-dimensional ($d=2$) and three-dimensional ($d=3$)
cases, respectively. In general, a superdisk possesses square 
symmetry, as $p$ moves away from unity,
two families of superdisks with square symmetry can be obtained,
with the symmetry axes rotated 45 degrees with respect to each
other; when $p<0.5$, the superdisk is concave (see Fig.~\ref{fig1}).

\begin{figure}
\begin{center}
$\begin{array}{c@{\hspace{0.75cm}}c@{\hspace{0.75cm}}c@{\hspace{0.75cm}}c}
\includegraphics[height=3.0cm, keepaspectratio]{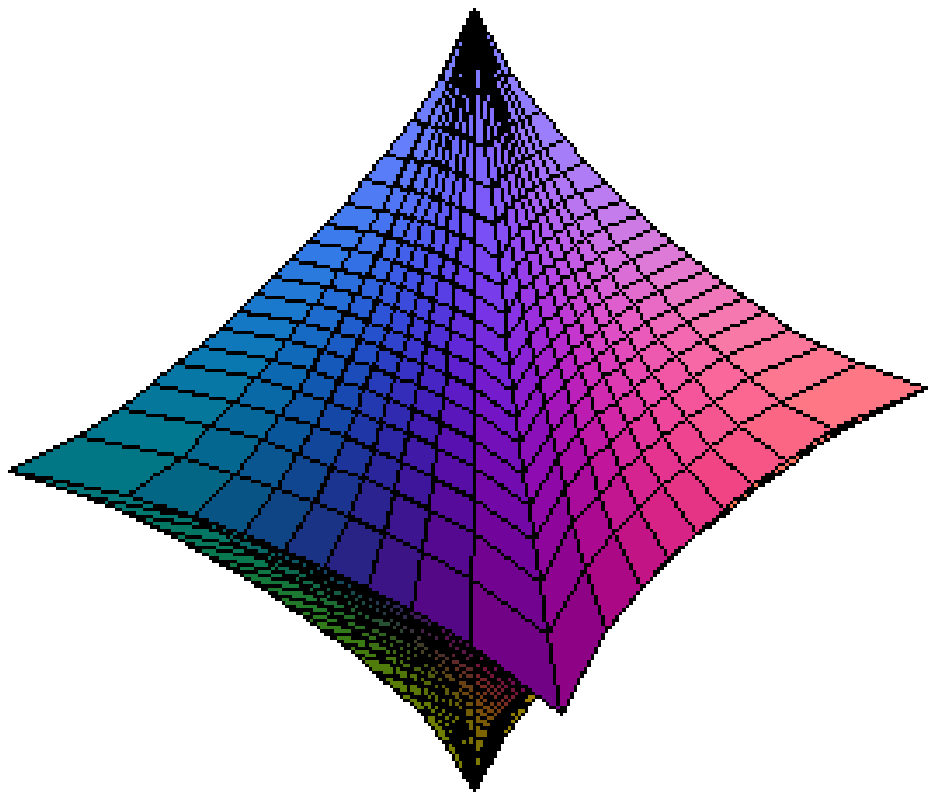} &
\includegraphics[height=3.0cm, keepaspectratio]{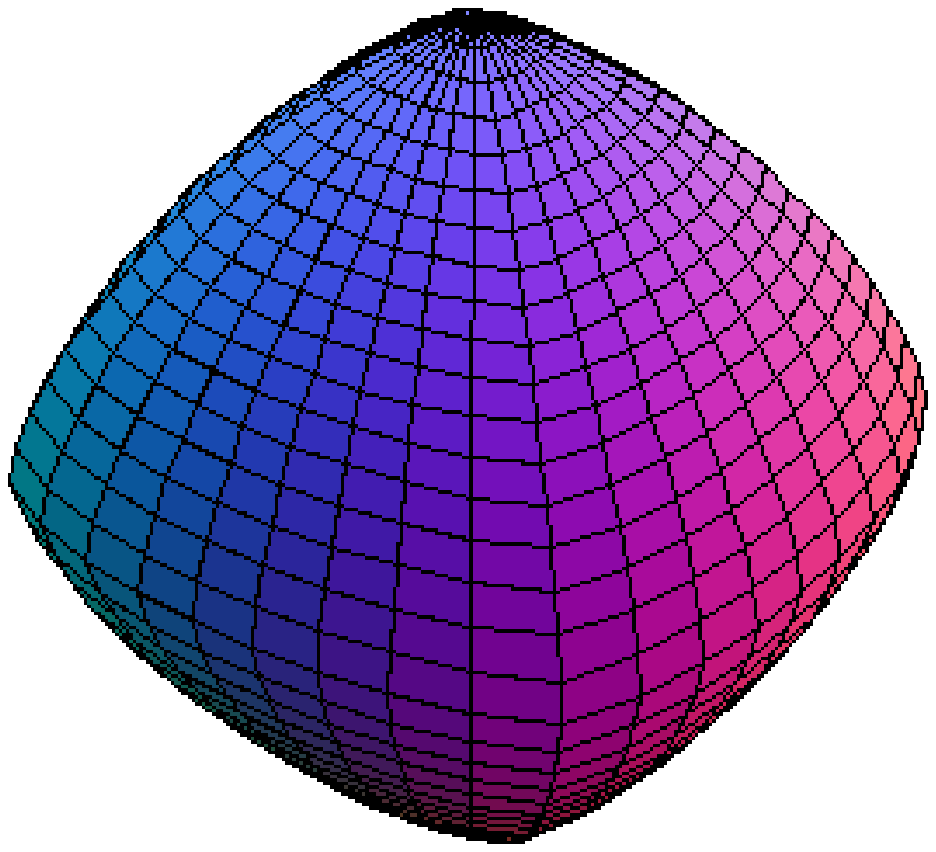} &
\includegraphics[height=3.0cm, keepaspectratio]{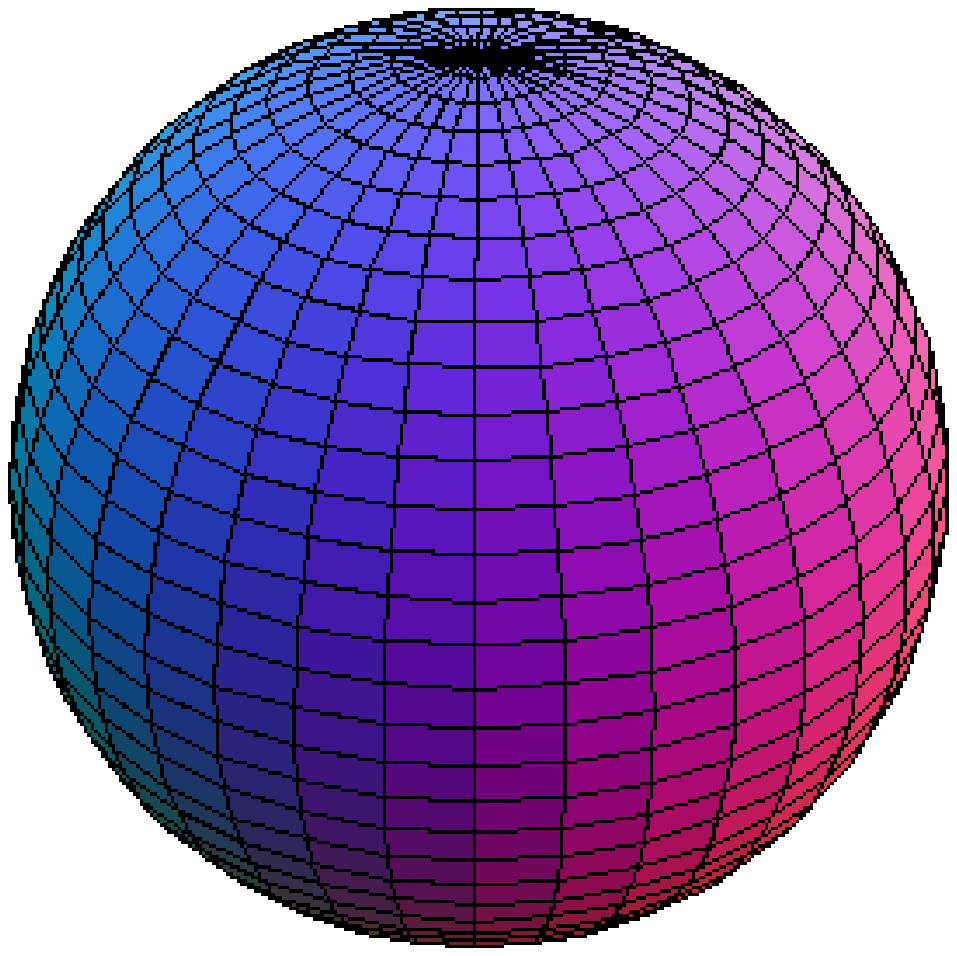} &
\includegraphics[height=3.0cm, keepaspectratio]{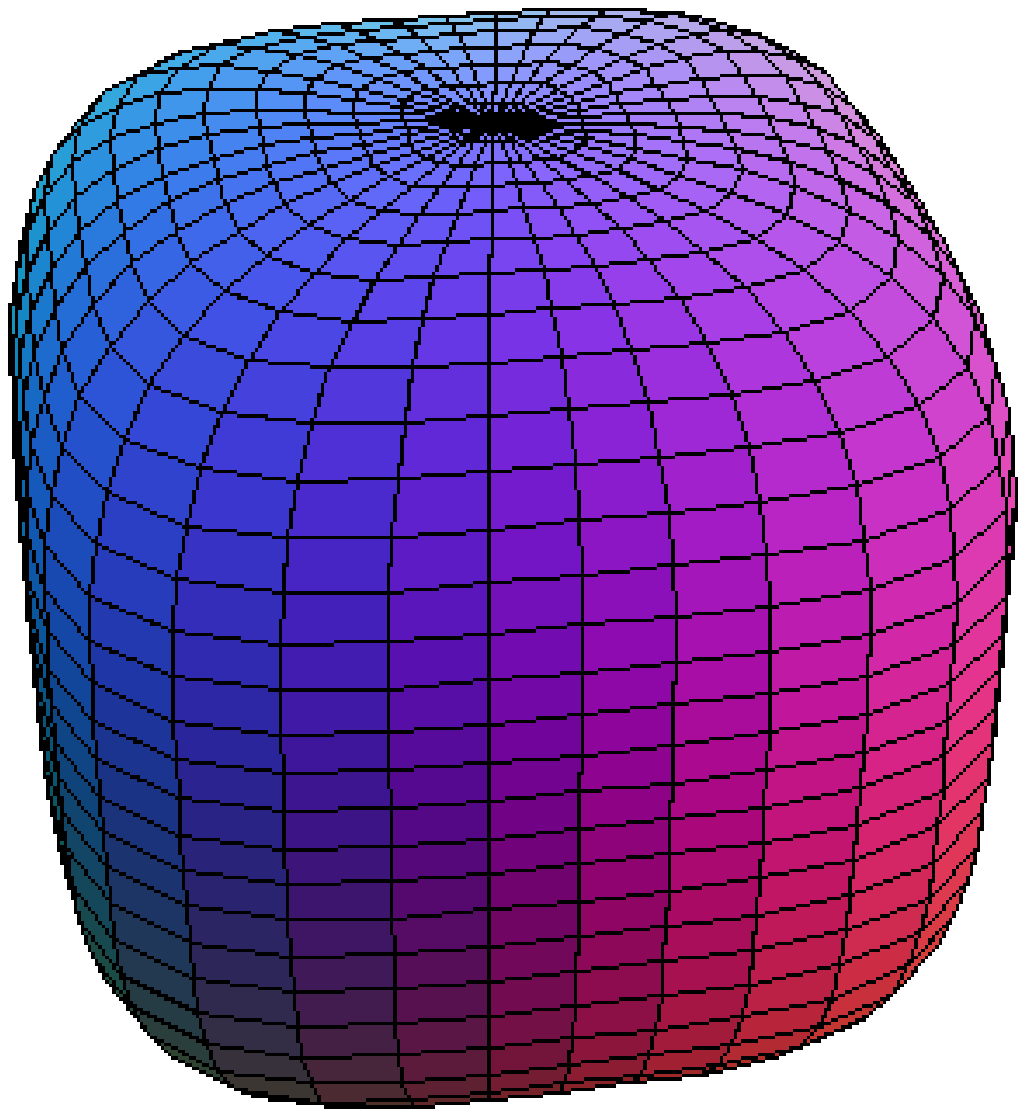} \\
\mbox{(a) p = 0.45} & \mbox{(b) p = 0.75} & \mbox{(c) p = 1.0} &
\mbox{(d) p = 2.0}
\end{array}$
\end{center}
\caption{(color online). Superballs with different values of the deformation
parameter $p$.} \label{fig2}
\end{figure}

In three dimensions, a superball is a perfect sphere at $p=1$, 
but can possess two types of shape anisotropy: cubic-like shapes 
(three-dimensional analog of the square symmetry of the superdisk) 
and octahedral-like shapes, depending on the value of the 
deformation parameter $p$ (see Fig.~\ref{fig2}). 
As $p$ continuously increases
from 1 to $\infty$, we have a family of convex superballs with
cubic-like shapes; at the limit $p = \infty$, the superball is a
perfect cube. As $p$ decreases from 1 to 0.5, a family of convex
superballs with octahedral-like shapes are obtained; at $p = 0.5$,
the superball becomes a regular octahedron. When $p<0.5$, the
superball still possesses an octahedral-like shape  but is now concave, 
becoming a three-dimensional \textit{``cross"} in the limit $p
\rightarrow 0$ \cite{shape}. Note that the cube and regular 
octahedron (two of the five Platonic polyhedra) have the same group 
symmetry (i.e., they have the same 48 space group elements) because they 
are dual to each other \cite{footnote1}.

Recently, we constructed the densest known packings of superdisks
in $\mathbb{R}^2$, which provides a wide class of packings of both
convex and concave particles with square symmetry
\cite{Superdisk}. In particular, the optimal packing is achieved
by one of two families of Bravais lattice packings, which we
called the $\Lambda_0$- and $\Lambda_1$-lattice packings. (Note that 
in a Bravais lattice packing there is only one particle in the 
fundamental cell in contrast to periodic packings that have multi-particle 
fundamental cells.) In the following, we will usually use the term
``lattice" to mean a ``Bravais lattice." As the
particle shape moves away from the circular disk point, the
packing density increases dramatically from $\pi/(2\sqrt 3)$ and
shows a non-analytic behavior at the circular point. Here we stress that in
light of a theorem due to Fejes T{\' o}th \cite{Fe64,Pach}, which states that the optimal
packing of any centrally symmetric convex body in $\mathbb{R}^2$
can be always achieved by circumscribing the body with the
smallest hexagon that tiles the plane, our packings can be
verified to be optimal for every specific value of $p>0.5$. This
procedure is outlined in the Appendix of this paper.

In this paper,  we construct the densest known packings of
convex superballs, which are suggested by simulations and based on
the requirements that certain group symmetries of the packings 
are preserved as the superballs deform from the sphere and octahedron points. 
Moreover, we show that as $p$ changes from
unity, i.e., moves away from the sphere point, 
maximal packing density $\phi_{max}$ rises steeply.
The broken rotational symmetry of superballs results in a cusp in
$\phi_{max}$ at $p=1$. Thus, the initial increase of $\phi_{max}$
is linear in $|p-1|$ and $\phi_{max}$ is a nonanalytic function of
$p$ at $p=1$. Two more  
nontrivial nonanalytic behaviors occur at $p^*_c = 1.1509\ldots$ and 
$p^*_o=\ln3/\ln4=0.7924\ldots$ for ``cubic'' and ``octahedral'' 
superballs, respectively,

For superballs in the cubic regime ($p>1$), the
candidate optimal packings are achieved by two families of Bravais
lattice packings possessing two-fold (see Sec. III.A) and three-fold rotational 
symmetry (see Sec. III.B), respectively, which
can both be considered to be continuous deformations of the FCC lattice.
For superballs in the octahedral regime ($0.5<p< 1$), there are also 
two families of Bravais lattices obtainable from continuous
deformations of the FCC lattice keeping its four-fold rotational
symmetry (see Sec. III.B), and from the densest lattice packing for regular octahedra
\cite{Mink,Henk}, keeping the translational symmetry of the projected
lattice on the coordinate planes, which are apparently optimal
in the vicinity of the sphere point and the octahedron point,
respectively (see Secs. IV.A and IV.B). The fact that the two families of lattice packings for 
superballs with octahedral shapes 
constructed separately from the provable optimal packing of spheres and the densest 
known packing of octahedra (which we conjecture to be optimal \cite{footnote2})
 meet at a single point (the same packing density with 
the same $p$ value) strongly suggests that our candidate packings are most likely  
optimal. The packing characteristics determined by the broken rotational
symmetry of superballs are similar to but richer than their
two-dimensional ``superdisk" counterparts \cite{Superdisk} 
and are distinctly different from that of ellipsoid  packings \cite{Alexks}.

For concave superballs ($p<0.5$), the lack of simulation techniques 
to generate such dense packings prevent us from drawing definitive conclusions
about the optimality of concave superball packings. Here we construct a family of dense packings of
concave superballs that are close to optimal around the octahedron
point ($p=0.5$), based on the densest Bravais lattice packing and will discuss
an interesting aligning effect in the limit $p\rightarrow 0$ (see Sec. IV.C).

The rest of the paper is organized as follows: In Sec.~II, we
briefly describe the simulation techniques used to generate dense
packings of superballs, which guide our analytical
packing constructions. In Sec.~III and Sec.~IV, we construct the
densest known packings of superballs in the cubic and octahedral
regimes, respectively. The structural characteristics of the packings and the affect
of the broken symmetry of the superballs are discussed in detail. In
Sec.~V, we present concluding remarks. In the Appendix, we 
briefly outline the procedure
to show that for every specific value of $p$ the constructed packings
of two-dimensional superdisks in Ref.~\cite{Superdisk} are indeed
optimal.

\section{Dense Packings of Superballs via Computer Simulations}

\subsection{Event-Driven Molecular Dynamics Algorithm for Superballs in the Cubic Regime}

Recently, Donev, Torquato and Stillinger developed a highly
efficient event-driven molecular dynamics packing algorithm
\cite{AleksII}, which generalizes the Lubachevsky-Stillinger (LS)
sphere-packing algorithm \cite{LSpacking} to the case of other
centrally symmetric convex bodies (e.g., ellipsoids and
superballs). Henceforth, we refer to the algorithm as the
Donev-Torquato-Stillinger (DTS) algorithm for convenience. The DTS
algorithm works as follows: Initially, small particles (in our
case the superballs) are randomly distributed and randomly
oriented in the simulation box (fundamental cell) with periodic
boundary conditions and without any overlap. The particles are
then given translational and rotational velocities randomly and
their motion followed as they collide elastically and also expand
uniformly, while the fundamental cell deforms to better
accommodate the configuration. After some time, a jammed state
with a diverging collision rate $\gamma$ is reached and the
density reaches a local maximum value.

\begin{figure}
$\begin{array}{c@{\hspace{1.5cm}}c}
\includegraphics[height=6.0cm, keepaspectratio]{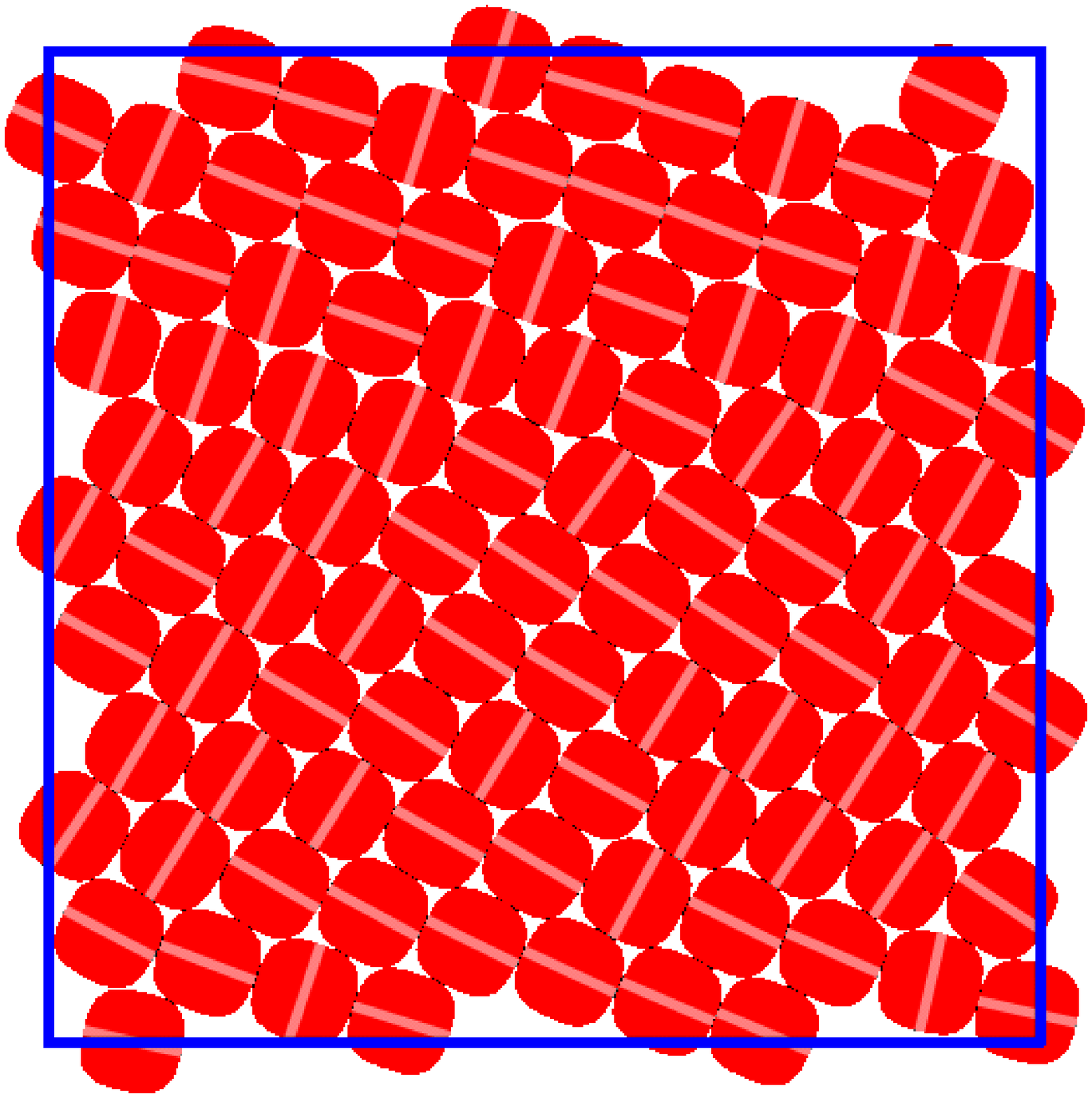} &
\includegraphics[height=6.0cm, keepaspectratio]{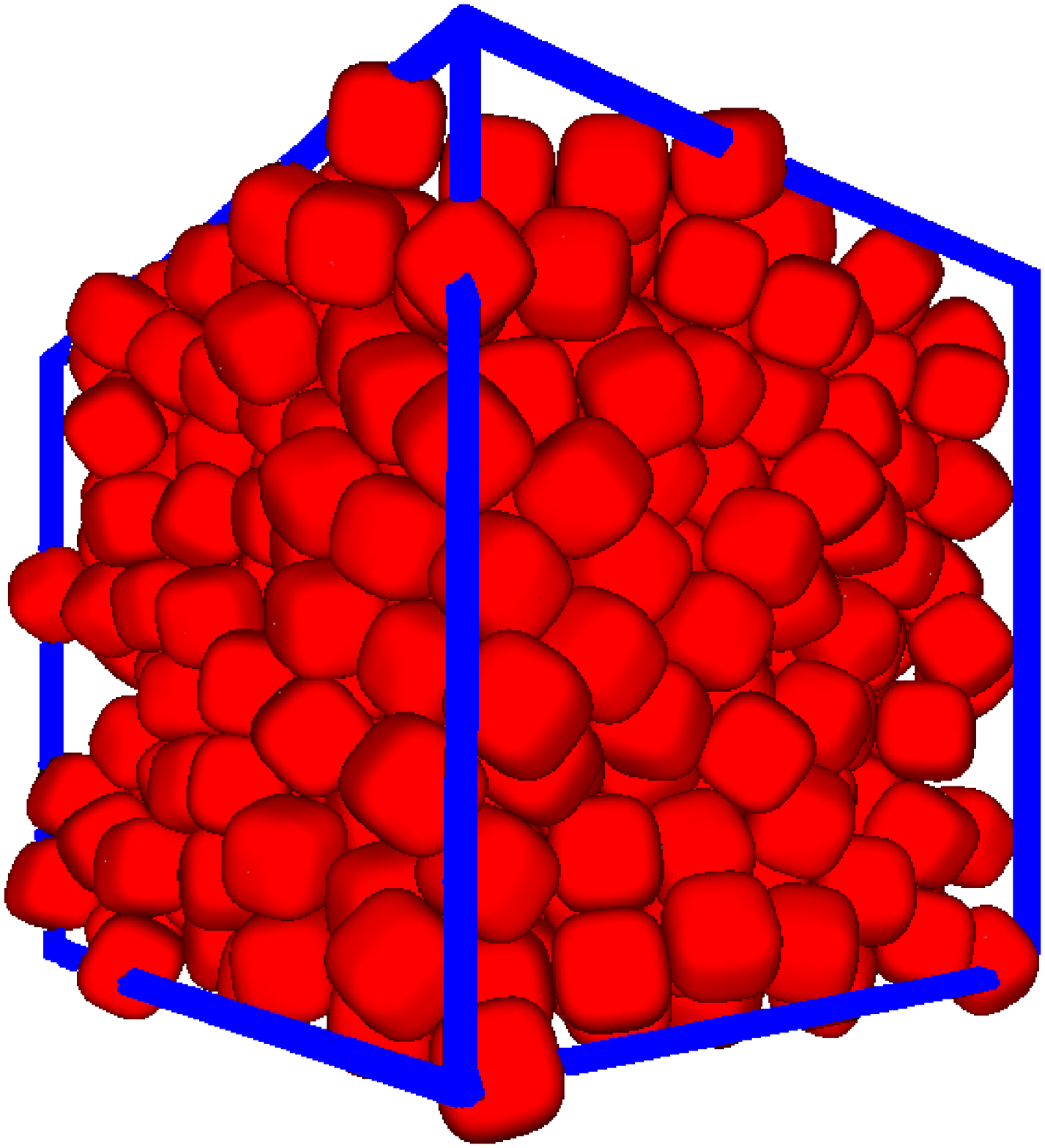} \\
\mbox{(a)} & \mbox{(b)}
\end{array}$
\caption{(color online). Dense packings of superdisks (a) and
superballs (b) with $p=1.5$ generated via the
Donev-Torquato-Stillinger algorithm with the growth rate $\gamma =
10^{-5}$. The packing densities are $\phi_a = 0.9123\ldots$ and
$\phi_b = 0.7072\ldots$, respectively. One can see that the
superdisk packing is nearly completely crystallized, while the
superball packing shows no significant signs of crystallization. The white
``chord'' in each superdisk indicates one of its symmetry axis.}
\label{fig3}
\end{figure}

\begin{figure}
\begin{center}
$\begin{array}{c@{\hspace{1.5cm}}c@{\hspace{1.5cm}}c}\\
\includegraphics[height=4.25cm, keepaspectratio]{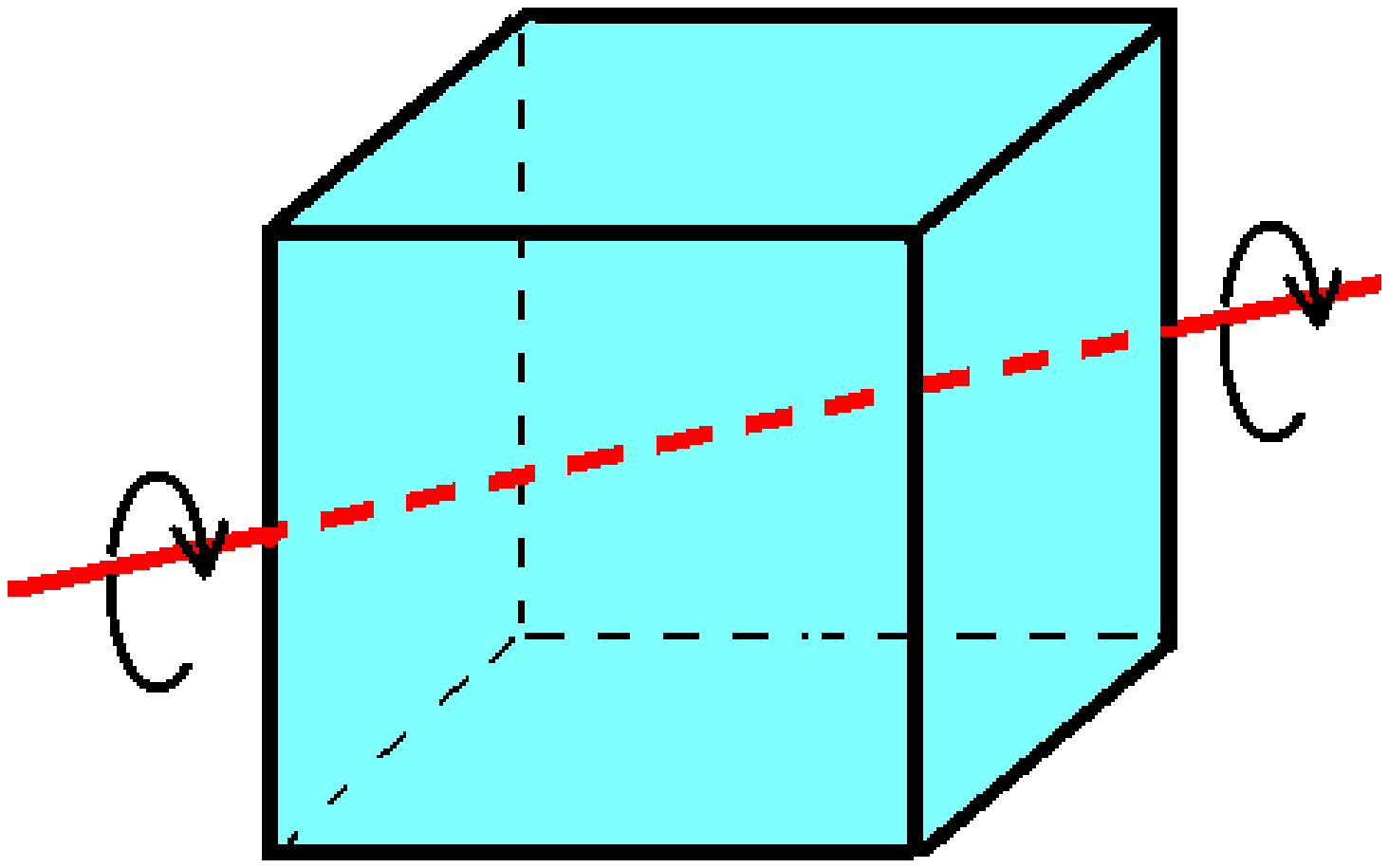}&
\includegraphics[height=4.5cm, keepaspectratio]{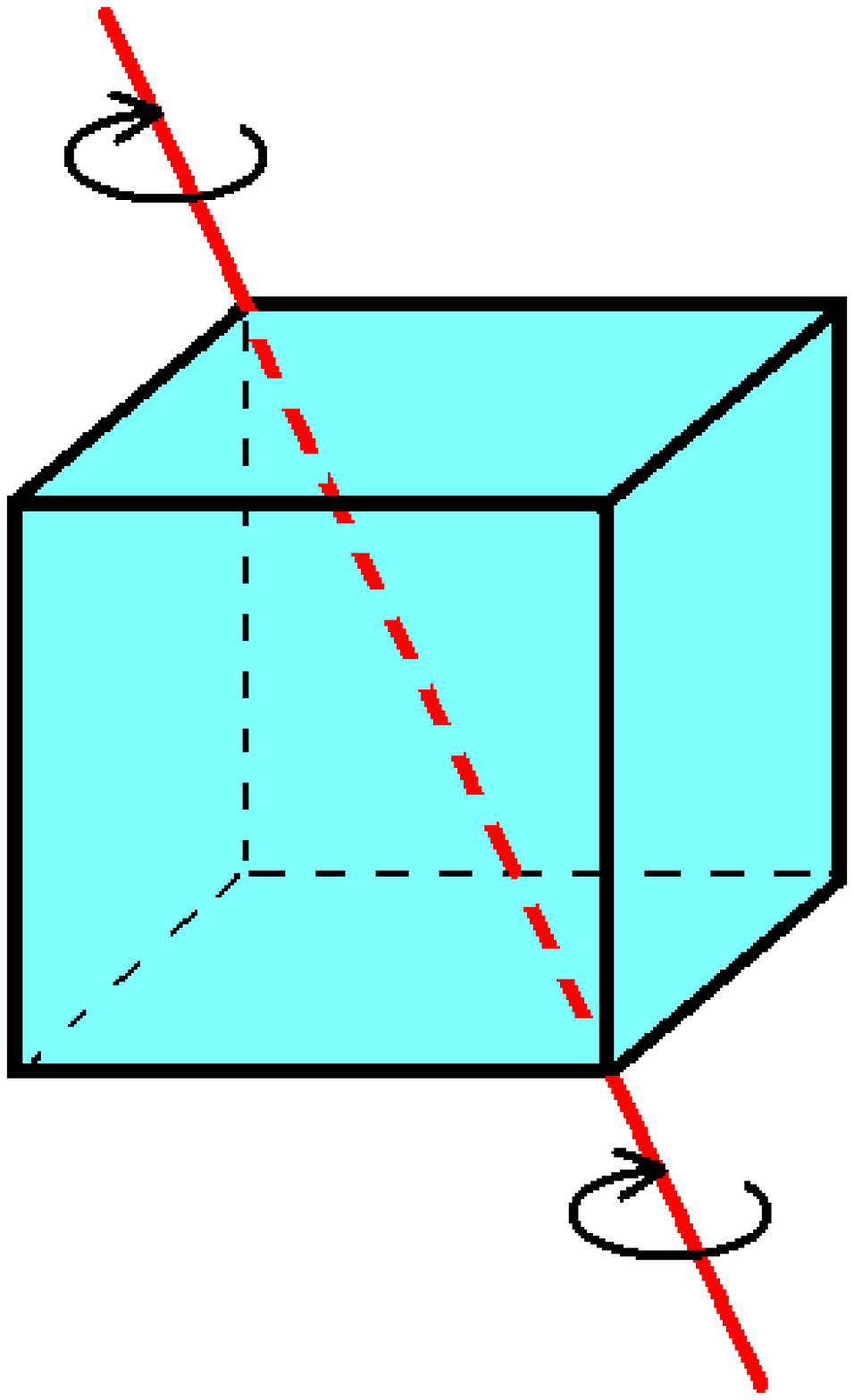}&
\includegraphics[height=4.5cm, keepaspectratio]{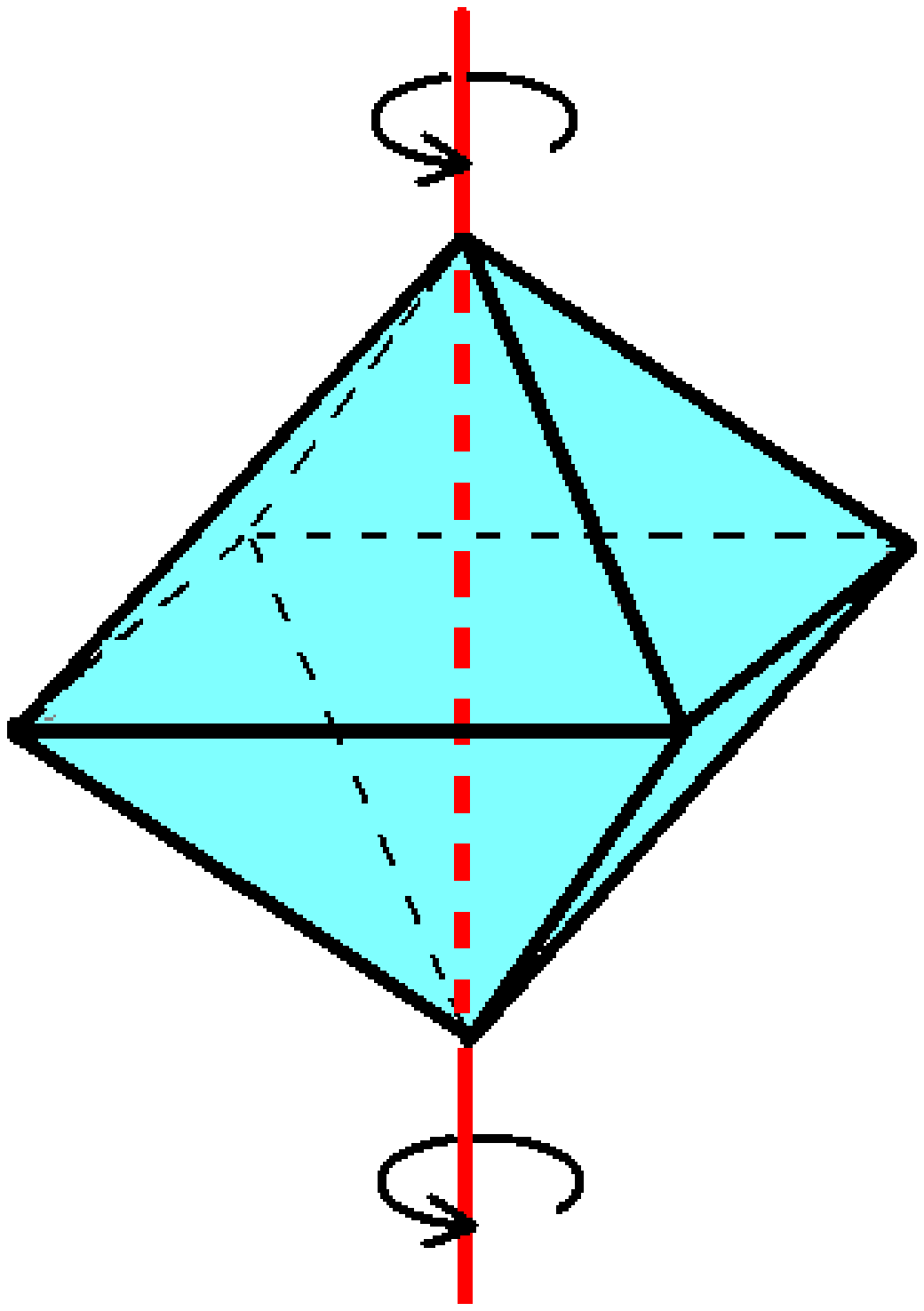}\\
\mbox{\bf (a)} & \mbox{\bf (b)}  & \mbox{\bf (c) }
\end{array}$
\end{center}
\caption{(color online). Illustrations of certain rotational
symmetries of the cube and regular octahedron. (a) Two-fold rotational symmetry
of a cube (axis shown coincides with a face diagonal). (b)  Three-fold rotational symmetry
of a cube (axis shown coincides with a body diagonal). (c) Four-fold rotational symmetry
of a regular octahedron. } \label{fig5.5}
\end{figure}

In order to generate the densest packing (or a packing close to
the densest one) of superballs in the cubic regime ($p>1$), a
sufficiently slow growth rate is necessary, as verified by the
extensive studies on spheres and circular disks, ellipsoids and
ellipses as well as superdisks \cite{SalPRL, Kansal, Alexks,
Superdisk}. In two dimensions, the densest local packing of many
centrally symmetric convex particles (among which circular disks,
ellipses and superdisks are of particular interest) can tessellate
the space. The consistency of local and global optima results in
nearly complete crystallization in large packings of these
particles [e.g., see Fig.~\ref{fig3}(a)]. By contrast, in three dimensions
the geometrical frustration of superballs (i.e., the densest local 
packing is not consistent with 
the optimal global packing) makes it very difficult
for the system to follow the equilibrium branch of the phase
diagram (without becoming stuck in any local minima) all the way
to the densest packing state, even when small growth rates are
used [e.g., see Fig.~\ref{fig3}(b)].

To resolve this difficulty, one could specify initial
configurations that are not random. A good initial
configuration would be unsaturated \cite{sat} lattice packings that
are hypothesized to be similar to the optimal lattice packing, 
which can be obtained by a reasonable guess. Here we
choose to arrange the superballs on the Bravais lattices that
provide dense packings of spheres, such as the face-centered cubic
lattice and its stacking variants as well as the body-centered
cubic lattice, with a four-fold rotational symmetry axis of a
superball that is parallel to a coordinate axis.

Two kinds of highly dense lattice packings of superballs emerge
from the simulations [see Fig.~\ref{fig5}(a) and (b)] starting from the unsaturated
face-centered cubic packing as the initial configuration. Subsequent
analytical calculations suggested by these simulation results lead
us to the exact construction of two families of Bravias lattice
packings, which we call the $\mathbb{C}_0$- and $\mathbb{C}_1$-lattice packings 
(so named because they both involve particles with cubic-like  shapes), whose
global symmetries are consistent with the two-fold [see Fig~\ref{fig5.5}(a)] and 
three-fold rotational symmetry [see Fig.~\ref{fig5.5}(b)] of a
superball of the packing, respectively. The structural and packing characteristics of
the $\mathbb{C}_0$- and $\mathbb{C}_1$-lattice packings will be discussed in detail in
Sec.~III. We emphasize here that we do not exclude the possibility
of the existence of denser periodic packings with a complex
particle basis, although we did not find any of these packings by
running the simulations with a small number of particles in the
fundamental cell which facilitates occurrence of denser periodic
packings if they existed \cite{Alexks}.

\subsection{Stochastic Optimization Algorithm for  Superballs in Octahedral Regime}

For convex superballs in the octahedral regime ($0.5<p<1$), the
prediction of the collision sequence of particles, an essential step in the DTS
algorithm, is numerically unstable when the particle shape
deviates too much from a sphere [see Fig.~\ref{fig2}(b)]
\cite{AleksThesis}. Here we use a novel stochastic optimization packing
algorithm \cite{OptPacking} to generate dense packings of convex
superballs with octahedral-like shapes. Starting with a given initial
configuration, the positions and orientations of the particles as
well as the lattice vectors are considered as ``design
variables'', on which the packing density is dependent. Various
optimization techniques could be used to search the
design-variable space for a point associated with a (local)
maximum of packing density close to the initial value, subject to
the nonoverlapping conditions. This process is repeated and the
packing density is gradually increased until a global maximum is
reached. For superballs, a stochastic optimization technique
is employed whereby random small trial moves in the
design-variable space are generated such any move is accepted if 
it leads to an increase of the packing density without
causing any overlaps, and is rejected otherwise.

The final density of the packing generated via the optimization
algorithm is sensitively dependent on the initial configuration.
Thus, appropriate choices of initial configurations are crucial to
the determination of the optimal packings. Based on the results of
packings of superdisks and superballs in the cubic regime, it is
reasonable to assume that the optimal packing lattice deforms
continuously as the shape of the superball changes, 
while the group symmetry of the lattice consistent with 
the four-fold rotational symmetry of superballs is preserved. 
Near the sphere point, we take advantage
of the cubic two-fold rotational symmetry of the FCC lattice by orienting the
superballs with their four-fold rotational symmetry axis parallel
to those of the lattice (see Fig.~\ref{fig9}). Near the octahedron point,
we use Minkowski's densest lattice packing for regular octahedra
\cite{Mink} (with the density $\phi_{M} = 18/19= 0.9473\ldots$)
and insert a shrunken superball into each octahedron in the
packing. Note that Minkowski's lattice packing is the densest known
packing of regular octahedra, which we conjecture to be optimal
among all packings of regular octahedra, as 
explained in \cite{footnote2}. The unsaturated packings so
constructed are used as initial configurations for the
optimization algorithm.

Two types of highly dense lattice packings emerge from the
simulations [see Fig.~\ref{fig5}(c) and (d)], which we call the $\mathbb{O}_0$-
and $\mathbb{O}_1$-lattice packings, respectively (so named
because the particles have octahedral-like shapes). The packings
 and their structural characteristics will be discussed in detail in
Sec.~IV.

\section{Optimal Packing of Superballs in the Cubic Regime}

\subsection{The $\mathbb{C}_0$-Lattice Packing}

\begin{figure}
$\begin{array}{c}
\includegraphics[height = 6.5cm, keepaspectratio]{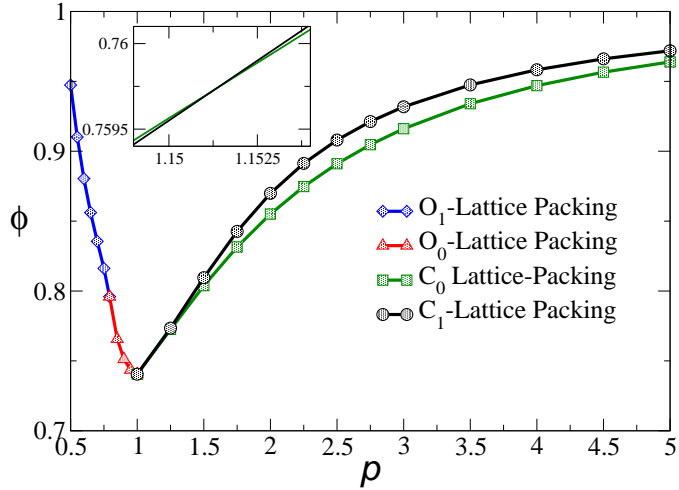}
\end{array}$
\caption{(color online). Density versus  deformation parameter $p$
for the packings of convex superballs. Insert: Around $p^*_c = 1.1509\ldots$, 
the two curves are almost locally parallel to each other.} \label{fig4}
\end{figure}

\begin{figure}
\begin{center}
$\begin{array}{c@{\hspace{0.2cm}}c@{\hspace{0.2cm}}c@{\hspace{0.2cm}}c}
\includegraphics[height=4cm, keepaspectratio]{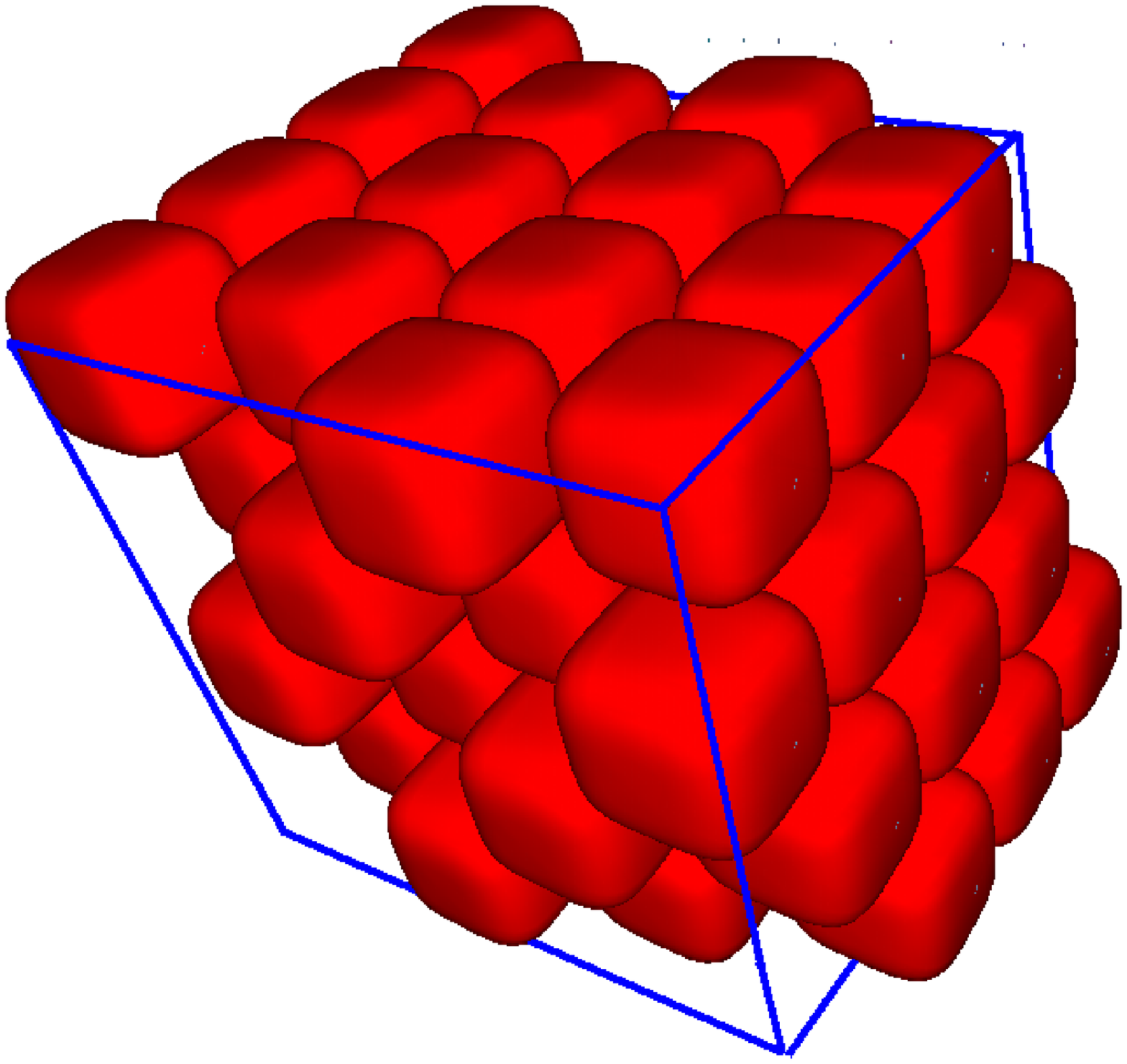} &
\includegraphics[height=4cm, keepaspectratio]{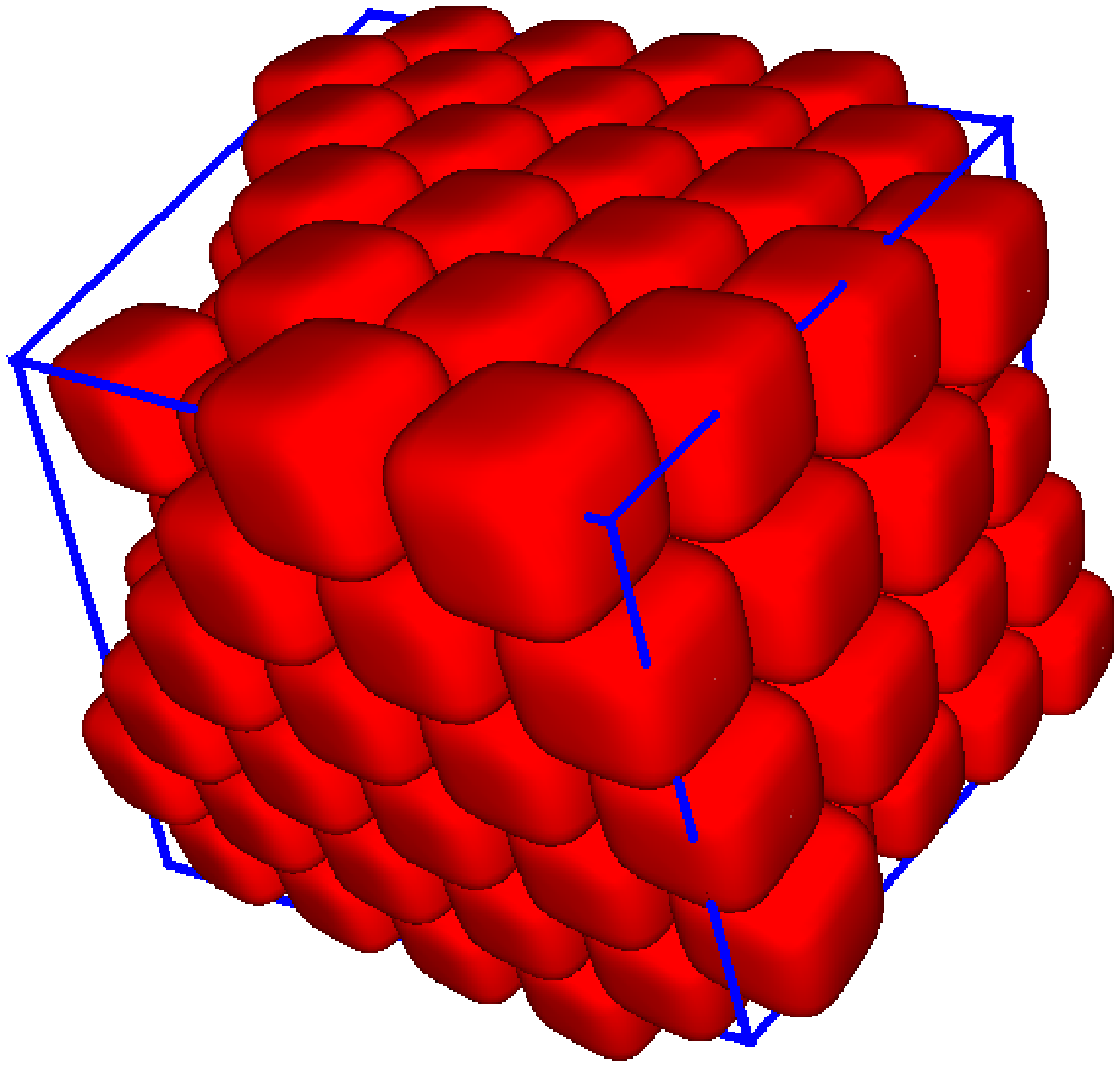} &
\includegraphics[height=4cm, keepaspectratio]{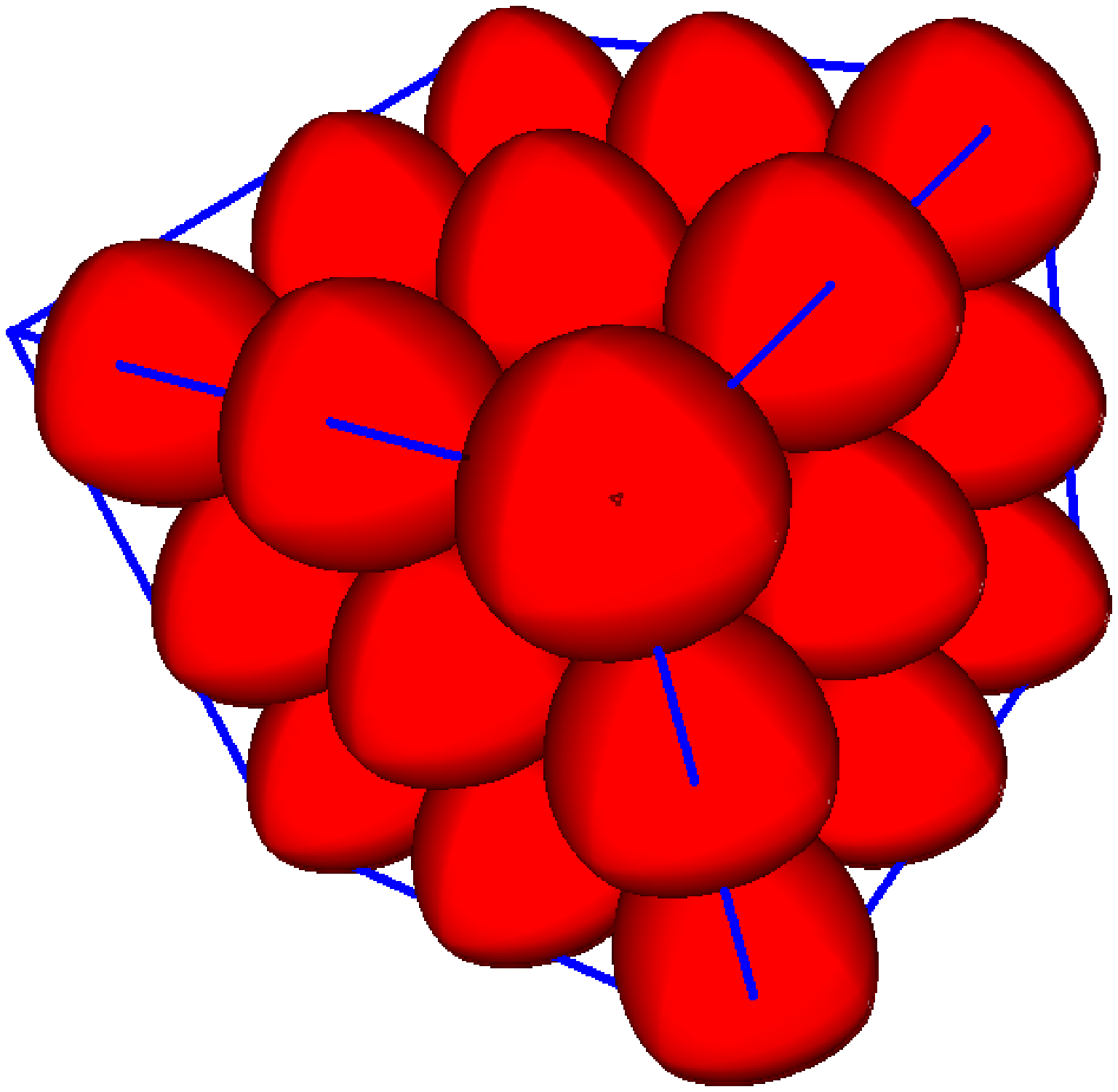} &
\includegraphics[height=4cm, keepaspectratio]{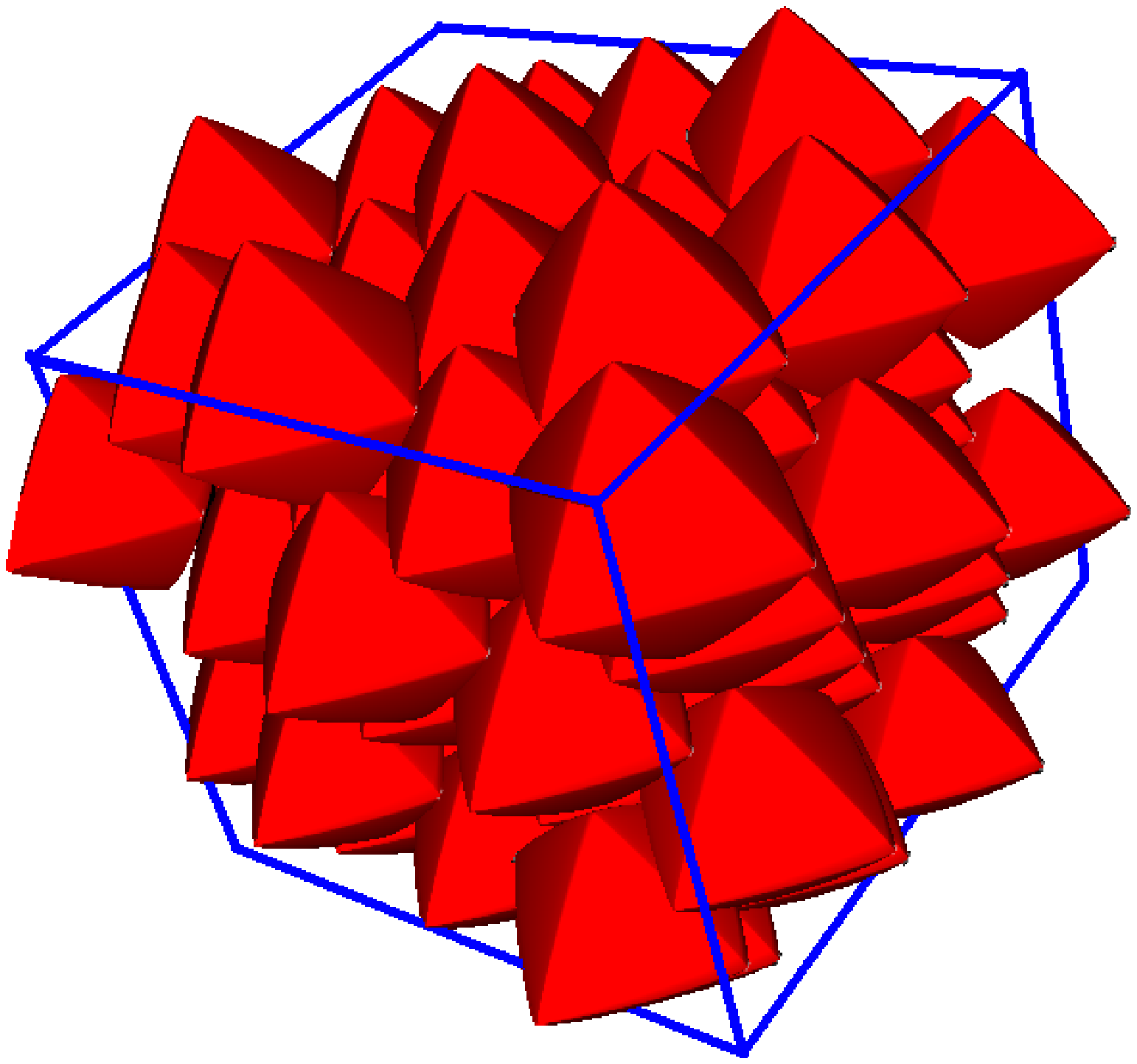} \\
\mbox{(a)} & \mbox{(b)} & \mbox{(c)}  & \mbox{(d)}
\end{array}$
\end{center}
\caption{(color online). Candidate optimal packings of superballs:
(a) The $\mathbb{C}_0$-lattice packing of superballs with $p = 1.8$. 
(b) The $\mathbb{C}_1$-lattice packing of superballs with $p = 2.0$. 
(c) The $\mathbb{O}_0$-lattice packing of superballs with $p=0.8$. 
(d) The $\mathbb{O}_1$-lattice packing of superballs with $p=0.55$.}
\label{fig5}
\end{figure}

\begin{figure}
\begin{center}
$\begin{array}{c@{\hspace{2.5cm}}c}\\
\includegraphics[height=4.5cm, keepaspectratio]{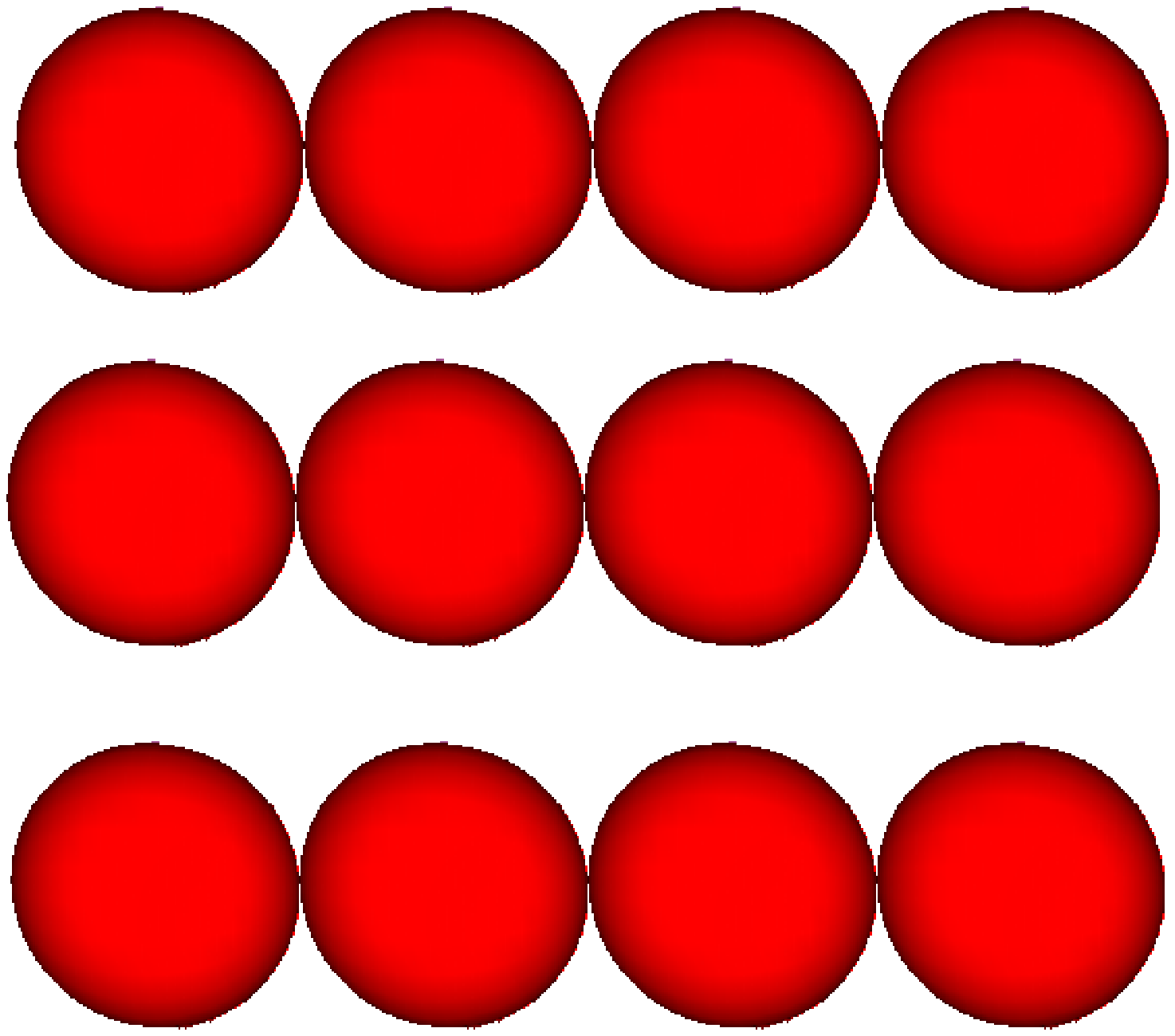}&
\includegraphics[height=4.5cm, keepaspectratio]{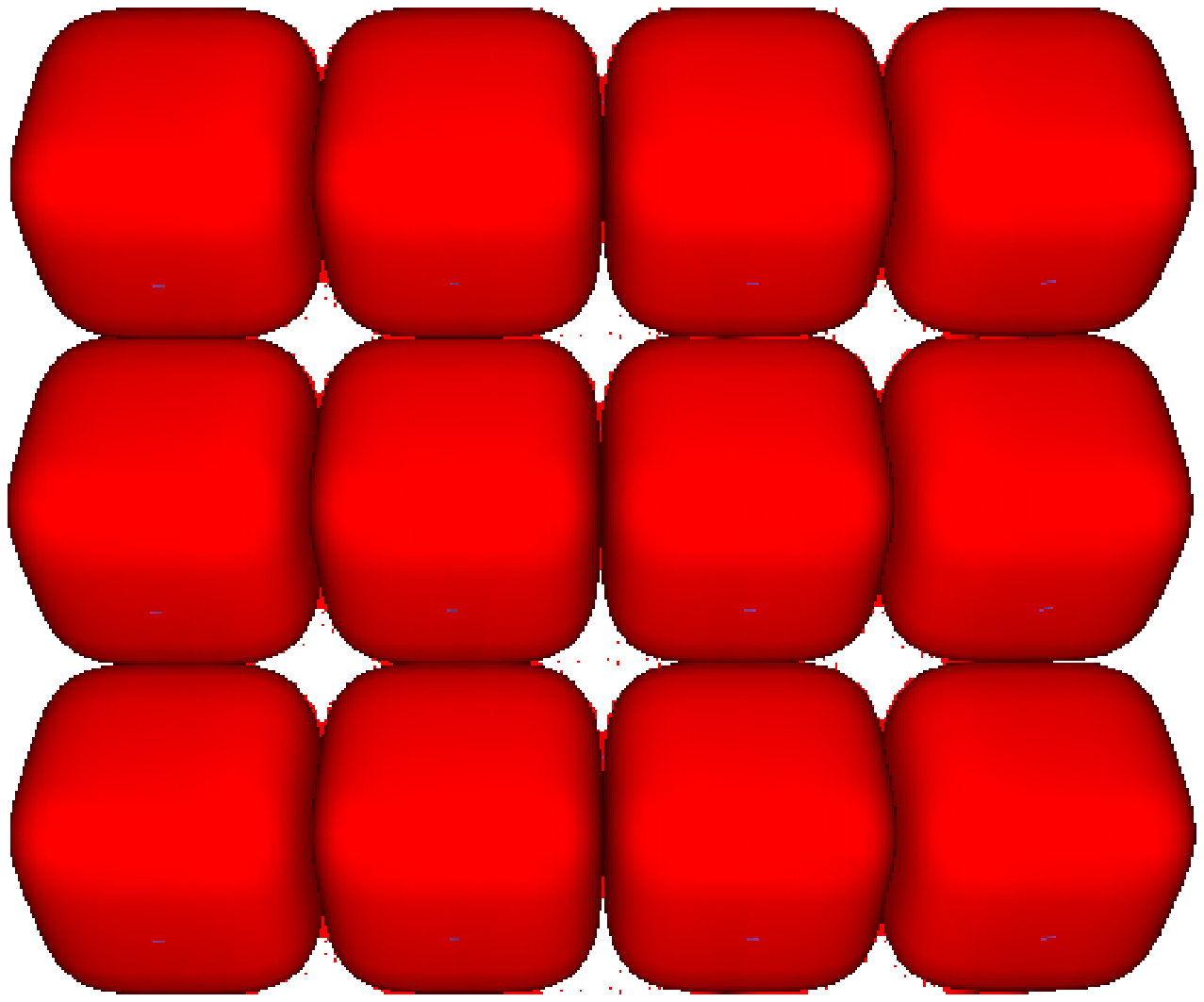}\\
\mbox{\bf (a)} & \mbox{\bf (b)}
\end{array}$
\end{center}
\caption{(color online). (a) The stretched square-lattice layer of spheres. 
(b) The stretched lattice layer of superballs.} \label{fig001}
\end{figure}

As suggested by the simulation results, our analysis shows that the densest packings of 
superballs with $1<p<p^*_c$ are given by the $\mathbb{C}_0$ lattices, where 
$p^*_c = 1.1509\ldots$ with $\phi^*_c = 0.7596\ldots$. Although the $\mathbb{C}_0$-lattice 
packings are only apparently optimal for $p \in (1, p^*_c)$, they exist for all $p\ge 1$, 
contrary to the $\mathbb{O}_0$- and $\mathbb{O}_1$-lattice packings of superballs with 
octahedral-like shapes, which only exist for certain range of $p$ as discussed below. 
The $\mathbb{C}_0$-lattice 
is obtainable from continuous deformation of the face-centered cubic lattice. In 
particular, the FCC lattice packing can be considered as a laminate of square-lattice 
planar layers [see Fig.~\ref{fig001}(a)]. The planar square-lattice must be stretched along 
one of its orthogonal directions so that the superballs in the cubic regime 
can be arranged on the (stretched) lattice site with one of its two-fold rotational 
symmetry axis along the stretched direction and the other one perpendicular to the 
plane of the square lattice, as shown in Fig.~\ref{fig001}(b). The 
magnitude of stretching is determined by the shape 
of the superballs, thus by the $p$ value. The stretched layers are then stacked 
so that the superballs in the top layers can sit exactly in the ``pockets'' formed 
by every four neighboring superballs in the bottom layer. Thus, each superball 
has 4 contacting neighbors in its own layers, 4 in the layer above and 4 in the layer 
below. The layer lamination is continued ad infinitum to construct the packing.

\begin{figure}
\begin{center}
$\begin{array}{c}\\
\includegraphics[width=5.5cm, keepaspectratio]{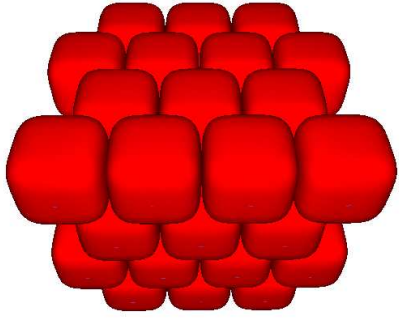}\\
\end{array}$
\end{center}
\caption{(color online). The 
$\mathbb{C}_0$ lattice for superballs in the cubic regime viewed from the
[110] direction. The structure remains invariant when it is rotated
$180^{\circ}$ about an axis in the [110] direction, showing its two-fold rotational symmetry.
} \label{fig002}
\end{figure}

The above construction uniquely defines the $\mathbb{C}_0$-lattice packings 
which possess two-fold rotational symmetry [see Fig.~\ref{fig002}] consistent with 
that of the superballs [see Fig.~\ref{fig5.5}(a)]. The lattice vectors 
for any $p$ value are given by

\begin{equation}
{\bf e}_1 = 2^{1-\frac{1}{2p}}{\bf i}+2^{1-\frac{1}{2p}}{\bf j},
\quad \displaystyle{{\bf e}_2 = 2{\bf k}}, \quad
\displaystyle{{\bf e}_3 = -2s{\bf i}+2(s+2^{-\frac{1}{2p}}){\bf
j}+{\bf k}},
\label{eq001}
\end{equation}

\noindent where ${\bf i}$, ${\bf j}$ and ${\bf k}$ are the unit
vectors along the coordinate axis and $s$ is the smallest positive root of the following
equation:

\begin{equation}
\displaystyle{\left ({s+2^{-\frac{1}{2p}}}\right
)^{2p}+s^{2p}+2^{-2p}-1 = 0.}
\label{eq002}
\end{equation}

The packing density $\phi_{max}$ is given by

\begin{equation}
\displaystyle{\phi = \frac{V_{sb}(p)}{|{\bf e}_1\times{\bf
e}_2\cdot{\bf e}_3|} = \frac{V_{sb}(p)}{2^{3-\frac{1}{2p}} \left
({2s+2^{-\frac{1}{2p}}}\right )}},
\label{eq003}
\end{equation}

\noindent where $V_{sb}(p)$ is the volume of superballs given by

\begin{equation}
\label{eq4} \displaystyle{V_{sb}(p) = \frac{2}{p^2}}B\left
({\frac{1}{2p},\frac{2p+1}{2p}}\right)B\left
({\frac{1}{2p},\frac{p+1}{p}}\right),
\end{equation}

\noindent and $B(x,y) = \Gamma(x)\Gamma(y)/\Gamma(x+y)$ and
$\Gamma(x)$ is the Euler Gamma function.

The packing density $\phi_{max}$ as a function of deformation
parameter $p$ is plotted in Fig.~\ref{fig4}. The
``right'' slope of $\phi_{max}$ at $p=1$ is given by

\begin{equation}
\displaystyle{ a_{+} = -\frac{\pi}{12\sqrt2}
\left[{{24+\ln 8+
4\Psi\left(\frac{1}{2}\right) + 2\Psi \left(\frac{3}{2}\right) -
6\Psi \left(\frac{5}{2}\right)} }\right]= 0.3555\ldots,}
\end{equation}

\noindent where $\Psi(z) = d[\ln \Gamma(z)]/dz$ is the digamma
function. The positive value of $a_+$ indicates that the initial
increase of the packing density is linear in $(p-1)$. As mentioned 
before, the $\mathbb{C}_0$-lattice packings are only optimal for 
$p \in (1, p^*_c)$, beyond $p^*_c$ the $\mathbb{C}_1$-lattice packings
 become the densest. For $p>1$ $\phi_{max}$
 increases dramatically until it reaches unity as the
particle shapes becomes more like a cube, which is more efficient
at filling space than a sphere. These characteristics stand in contrast to
those of the densest known ellipsoid packings, achieved by certain crystal
arrangements of congruent spheroids with a two-particle basis,
whose packing density as a function of aspect ratios has zero
initial slope and is bounded from above by a value of
$0.7707\ldots$ \cite{Alexks}. As we will see in Sec.~IV, as $p$
decreases from unity, the initial increase of $\phi_{max}$ is
linear in $(1-p)$. Thus, $\phi_{max}$ is a nonanalytic function of
$p$ at $p=1$, which is consistent with our conclusions about
superdisk packings. However it is distinctly different from the optimal spheroid
packings, for which $\phi_{max}$ increases smoothly as the aspect
ratios of the semi-axes vary from unity and hence has no cusp at
the sphere point \cite{Alexks}. The density of congruent ellipsoid
packings (not $\phi_{max}$) has a cusp-like behavior at the sphere
point only when the packings are randomly jammed \cite{AlexksIII}.
The distinction between the two systems results from different
broken rotational symmetries. For spheroids, the continuous
rotational symmetry is only partially broken, i.e., spheroids
still possess one rotationally symmetric axis; and the three
coordinate directions are not equivalent which facilitates dense
non-Bravais packings. For superballs, the continuous rotational
symmetry of a sphere is completely broken and the three coordinate
directions are equivalently four-fold rotationally symmetric
directions of the particle. Thus, a superball is less symmetric
but more isotropic than an ellipsoid which apparently prefers
dense Bravais lattice packings. The broken symmetry of superballs
makes their shapes more efficient in tiling space and thus results
in a larger and faster increase in the packing density. 

At $p = p^*_c = 1.1509\ldots$, the two lattice packings 
have the same density $\phi^*_c = 0.7596\ldots$ and superballs with 
$p^*_c$ possess a two-fold degenerate crystalline maximal density state. 
The $\mathbb{C}_0$ and $\mathbb{C}_1$ lattices have distinct group 
symmetries, thus the optimal packing ``jumps'' from one structure to 
another leading to another nonanalytic point in $\phi_{max}$. This 
behavior is similar with that of optimal superdisk packings.

\subsection{The $\mathbb{C}_1$-Lattice Packing}

\begin{figure}
\begin{center}
$\begin{array}{c@{\hspace{1.5cm}}c}\\
\includegraphics[height=4.5cm, keepaspectratio]{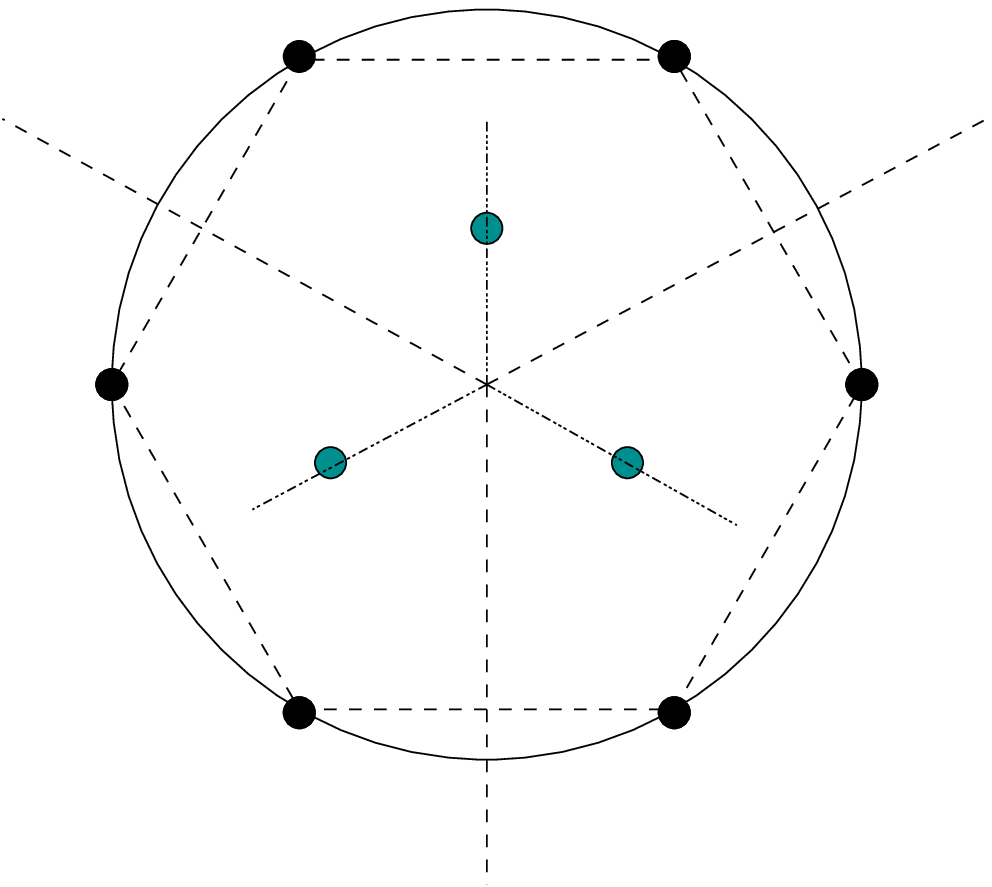}&
\includegraphics[height=4.5cm, keepaspectratio]{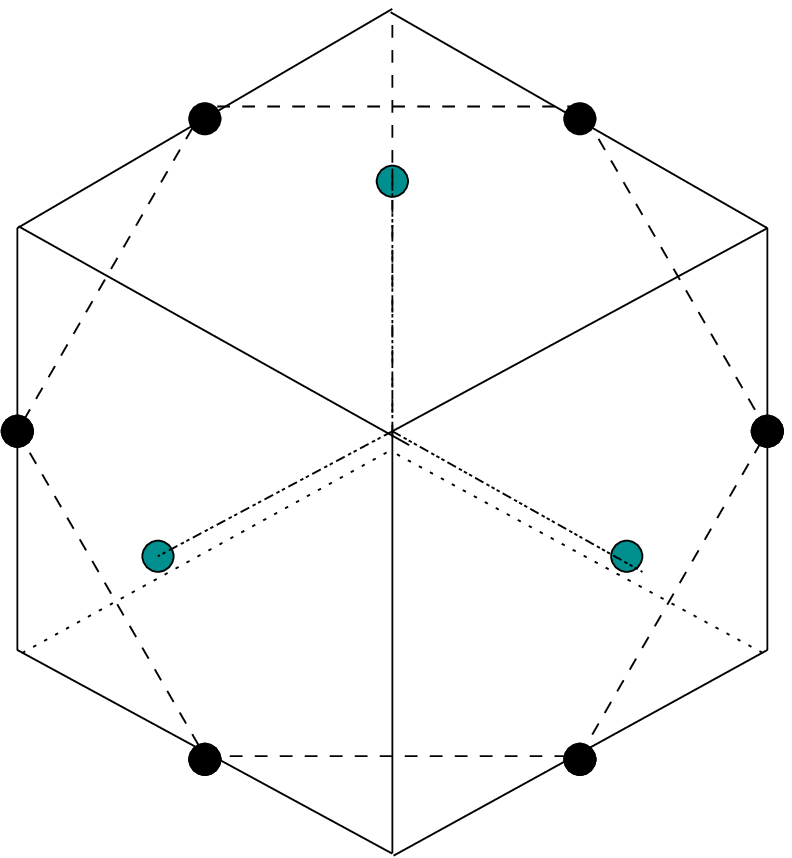}\\
\mbox{\bf (a)} & \mbox{\bf (b)}
\end{array}$
\end{center}
\caption{(color online). Contacting points of the a sphere (a) and
a schematic superball (shown as a cube) (b). The view is along the
[111] direction. The surface contacting points (blue) will
continuously move along the surface diagonals as the $p$ value
changes from unity to infinity. Every contacting point has a
symmetric image about the center of the superball and their
distributions have three-fold rotational symmetry.} \label{fig6}
\end{figure}

\begin{figure}
\begin{center}
$\begin{array}{c}\\
\includegraphics[width=5.5cm, keepaspectratio]{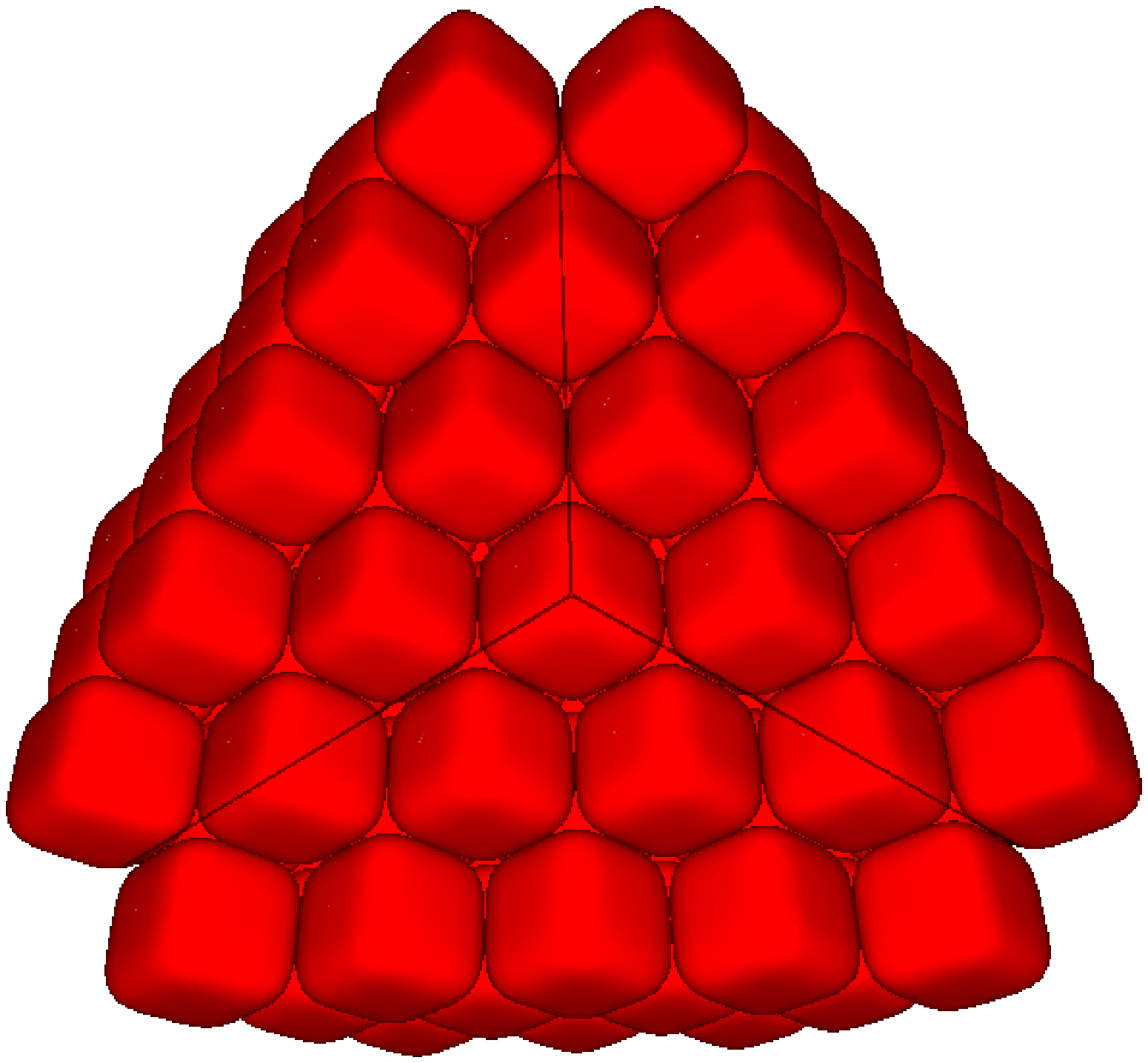}\\
\end{array}$
\end{center}
\caption{(color online). The 
$\mathbb{C}_1$ lattice for superballs in the cubic regime viewed from the
[111] direction. The thin dark lines show the three orthogonal
directions, which clearly exhibits the 3-fold rotational symmetry of the structure.} \label{fig7}
\end{figure}

For superballs with $p \ge p^*_c$ the densest packings 
are given by
the $\mathbb{C}_1$ lattices with three-fold rotational symmetry
consistent with that of the superballs, see Fig.~\ref{fig5}(b) and Fig.~\ref{fig7}. 
Note though the $\mathbb{C}_1$-lattice packings are only 
apparently optimal when $p\ge p^*_c$, they also exist for all $p \ge 1$. 
In the packing, each superball has 12 contacting neighbors, among which 6
contacting points are the edge centers and the other 6 are along
the surface diagonals, as schematically shown in Fig.~\ref{fig6}. At $p=1$,
this local packing (the central particle and its 12 neighbors)
possessing three-fold rotational symmetry is just the building
block of the face-centered cubic lattice packing of spheres. As
$p$ increases from unity, the continuous rotational symmetry of a
sphere is broken while the three-fold rotational symmetry of the
local packing could still be maintained since ``cubic'' superballs
still possess three-fold rotational symmetry [see Fig.~\ref{fig5.5}(b)]. Although the edge
contacting points will remain at the centers of the edges, the
surface contacting points will continuously follow a path along
the surface diagonals as $p$ changes from unity to infinity. As
$p$ approaches infinity, the surface contacting points will
approach the surface centers; when the superballs becomes a
perfect cube at $p=\infty$, the contacting points are degenerate,
i.e., the entire edges and surfaces of neighboring particles
contact each other, respectively.

The three-fold rotational symmetry of the local packing together
with the requirement that each particle has 12 contacting neighbors
uniquely determine the global packing lattice. And the packing
lattice for a specific $p$ value can be obtained by continuously
deforming the face-centered cubic lattice. In particular, the
lattice vectors of the $\mathbb{C}_1$ are given by

\begin{equation}
\label{eq1}
{\bf e}_1 = 2^{1-\frac{1}{2p}}{\bf i}+2^{1-\frac{1}{2p}}{\bf j},
\quad {\bf e}_2 = 2^{1-\frac{1}{2p}}{\bf i}+2^{1-\frac{1}{2p}}{\bf
k}, \quad \displaystyle{{\bf e}_3 = 2(s+2^{-\frac{1}{2p}}){\bf
i}-2s{\bf j}-2s {\bf k}},
\end{equation}

\noindent where $s$ is the smallest positive
root of the following equation:

\begin{equation}
\label{eq2} \displaystyle{\left ({s+2^{-\frac{1}{2p}}}\right
)^{2p}+2s^{2p}-1 = 0.}
\end{equation}

\noindent The packing density is readily computed from

\begin{equation}
\label{eq3} \displaystyle{\phi_{max} = \frac{V_{sb}(p)}{|{\bf
e}_1\times{\bf e}_2\cdot{\bf e}_3|} =
\frac{V_{sb}(p)}{2^{3-\frac{1}{p}} \left
({3s+2^{-\frac{1}{2p}}}\right )}},
\end{equation}

\noindent where $V_{sb}(p)$ is the volume of a superball with
deformation parameter $p$ as given by Eq.~(\ref{eq4}). The 
packing density $\phi_{max}$ is shown in Fig.~\ref{fig4}. 


\begin{figure}
\begin{center}
$\begin{array}{c@{\hspace{0.75cm}}c@{\hspace{0.75cm}}c}
\includegraphics[height=4.5cm, keepaspectratio]{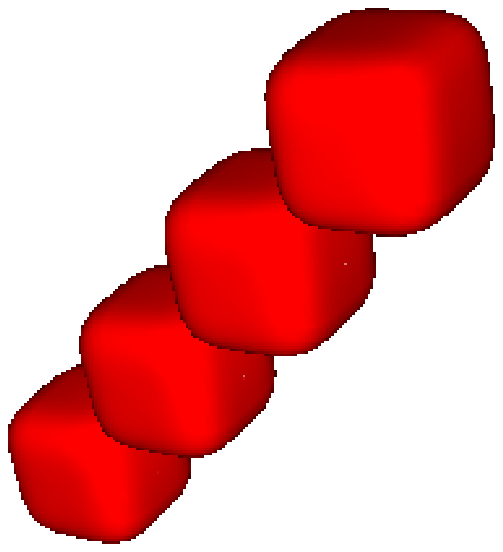} &
\includegraphics[height=4.5cm, keepaspectratio]{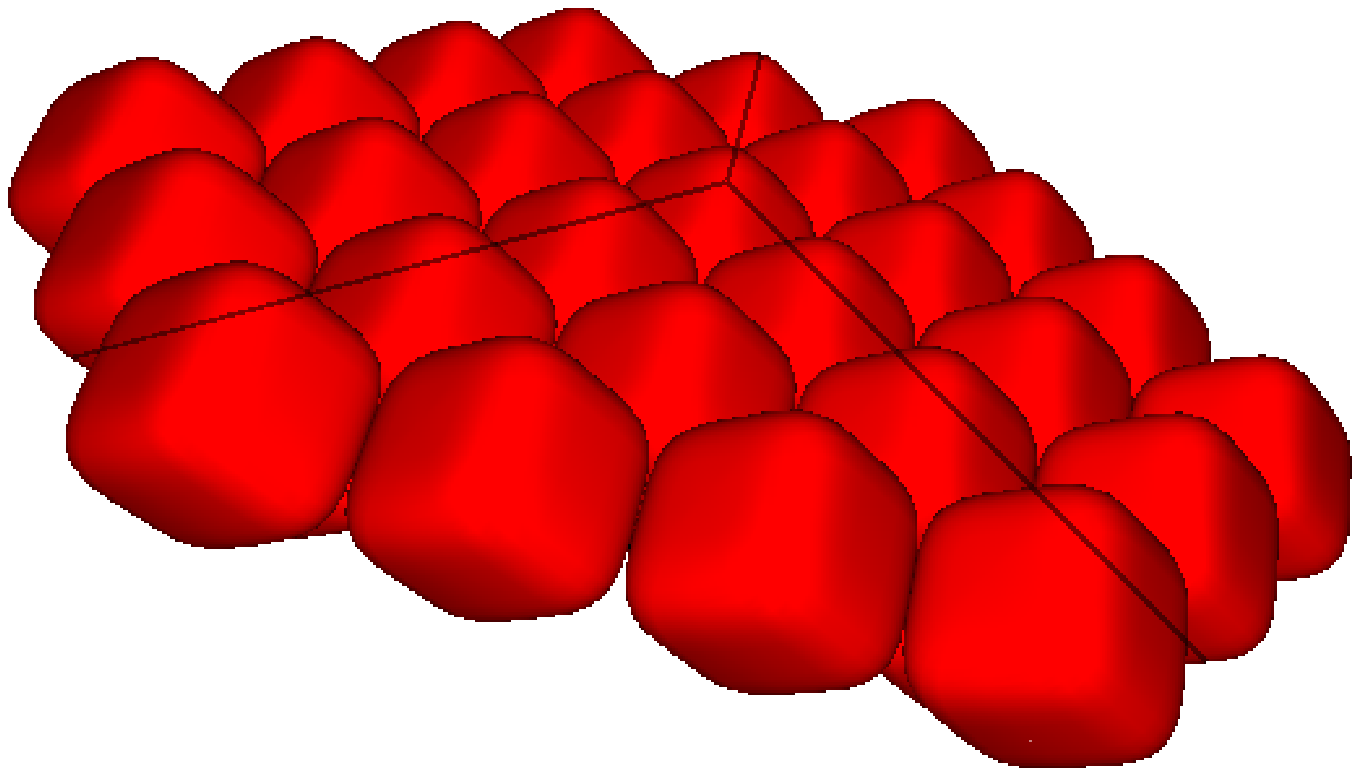} &
\includegraphics[height=4.5cm, keepaspectratio]{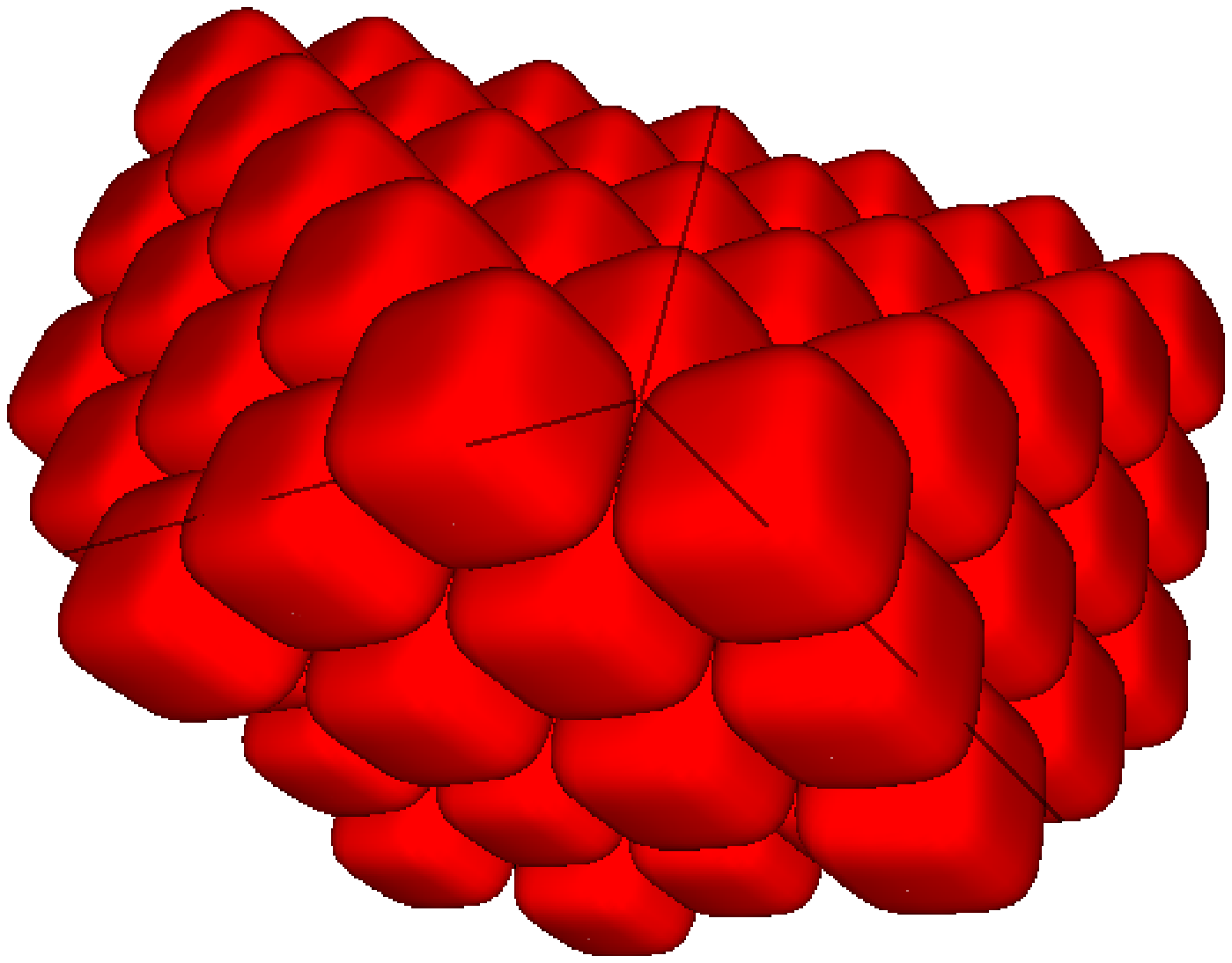} \\
\mbox{(a)} & \mbox{(b)} & \mbox{(c)}
\end{array}$
\end{center}
\caption{(color online). The $\mathbb{C}_1$-lattice packing of
superballs obtained by stacking superball chains: (a) The chain as
depicted in the text. (b) The layer constructed by stacking the
chains. (c) The packing obtained by stacking the layers.} \label{fig8} 
\end{figure}


As shown in Fig.~\ref{fig8}(a), a sequence of superballs contacting each
other through the centers of edges can be considered as a chain
which serves as the basic building block of the packing structure.
In contrast to the FCC packing of spheres in which the dense
layers constructed by stacking the chains of spheres are parallel
to one coordinate plane; for superballs with $p\neq1$, the chains
are shifted relative to each other and the normal direction of the
layer containing the centers of superballs does not coincide with
the coordinate directions. In this way, a lattice of ``pockets''
is formed in which the particles in the layer above can be exactly
fitted to construct the $\mathbb{C}_1$-lattice packing, as
illustrated in Fig.~\ref{fig8} (b) and (c).

\section{Optimal Packings of Superballs in the Octahedral Regime}

\subsection{The $\mathbb{O}_0$-Lattice Packing}

\begin{figure}
\begin{center}
$\begin{array}{c@{\hspace{2.5cm}}c}\\
\includegraphics[height=4.5cm, keepaspectratio]{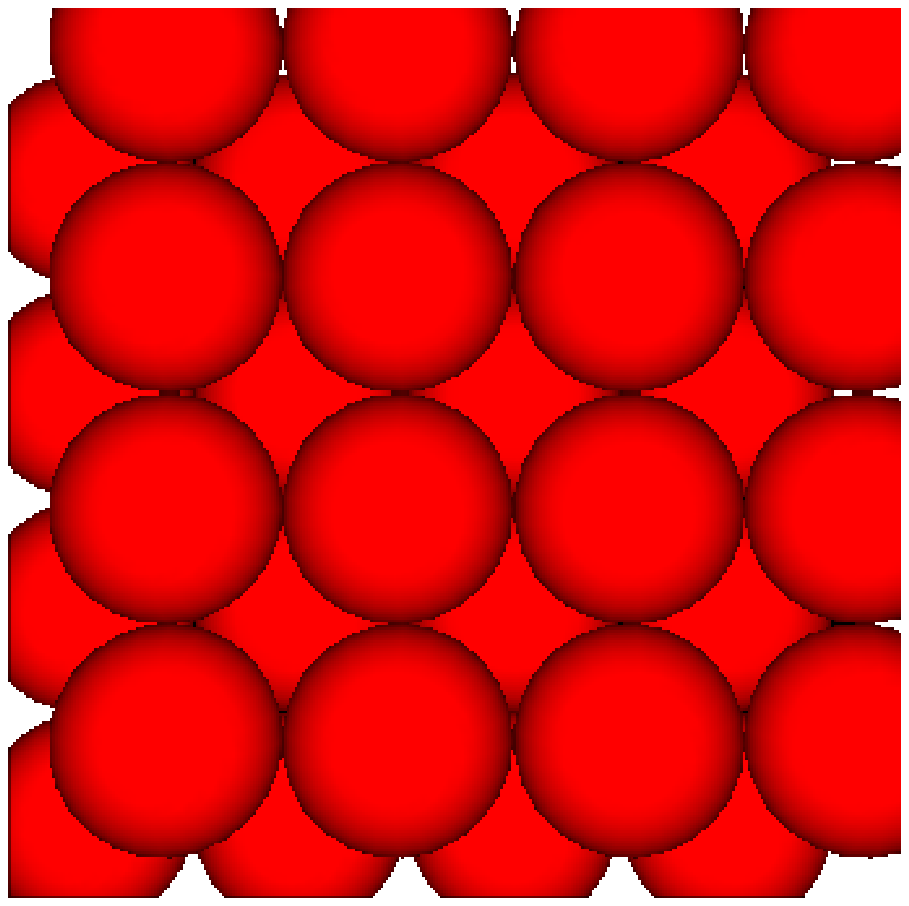}&
\includegraphics[height=4.5cm, keepaspectratio]{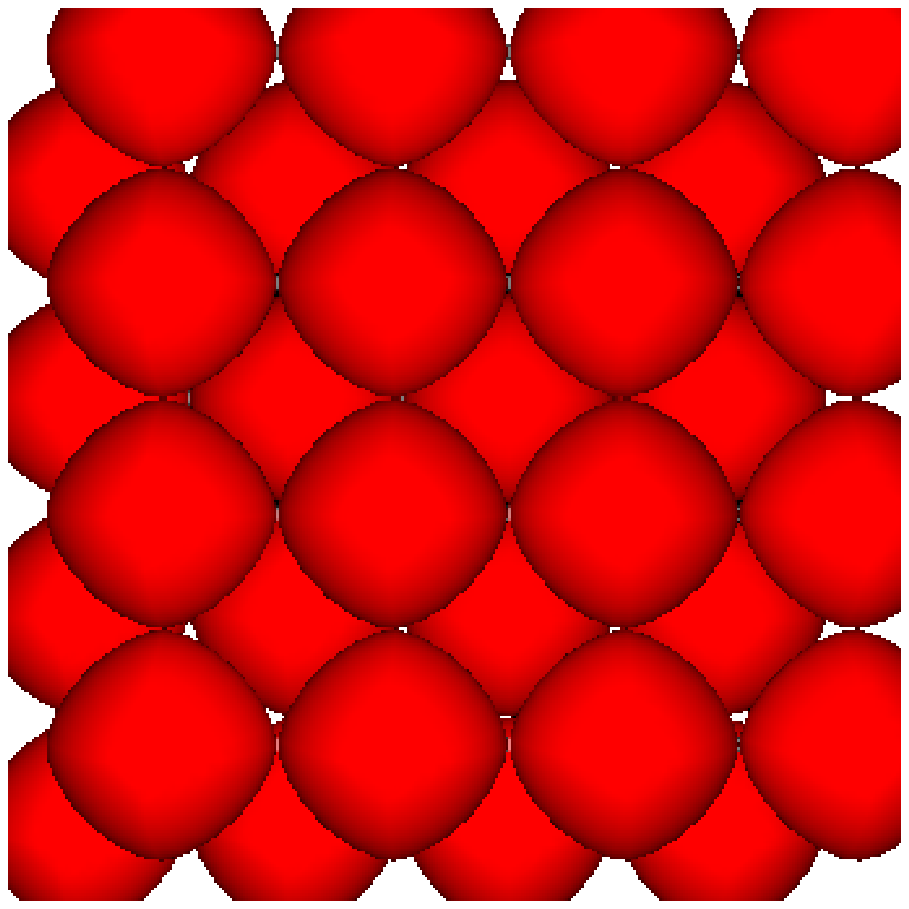}\\
\mbox{\bf (a)} & \mbox{\bf (b)}
\end{array}$
\end{center}
\caption{(color online). (a) Face-centered cubic packing of
spheres, viewed as a laminate of face-centered square layers from
the [001] direction. (b) Similar laminate of face-centered square
planar layers of superballs in the octahedral regime, viewed from
the [001] direction.} \label{fig9}
\end{figure}

As $p$ decreases from unity, the continuous rotational symmetry of
a sphere is broken and a superball has an octahedral-like shape
[see Fig.~\ref{fig2}(b)]. The $\mathbb{O}_0$ lattice [see Fig.~\ref{fig5}(c)] 
also can be
considered as a continuous deformation of the face-centered cubic
lattice. As before, we start from the FCC packing , viewed as a laminate of
square-lattice planar layers of spheres, as illustrated in
Fig.~\ref{fig9}(a). We similarly construct layers of superballs by
orienting one of the four-fold rotationally symmetric axis along
the normal of the layer [see Fig.~\ref{fig5.5}(c)], and aligning the other two of each
superball so that the square symmetry of cross-sections of
superballs in the layer is consistent with that of the layer, as
shown in Fig.~\ref{fig9}(b). The layers are stacked so that the superballs
in the top layers can exactly sit in the holes formed in the
bottom layer. The layer lamination is then continued ad infinitum
to generate the packing.

In the $\mathbb{O}_0$-lattice packing, each superball has 4 contacting
neighbors in its own layer, 4 in the layer above and 4 in the
layer below. The lattice vectors for any specific $p$ can be
uniquely determined by the symmetry and contacting requirements.
i.e.,

\begin{equation}
\label{eq5}
{\bf e}_1 = 2{\bf i}, \quad\quad {\bf e}_2 = 2{\bf j}, \quad\quad
\displaystyle{{\bf e}_3 = {\bf i}+{\bf j}+2\left
({1-2^{1-2p}}\right )^{\frac{1}{2p}} {\bf k}}.
\end{equation}
As the shape of a superball deviates more from a sphere,
the holes in the layer become larger and the distance between
successive layers becomes smaller. In the limit that the
superballs become octahedra ($p=0.5$), the planes of two successive
layers coincide with each other. However, the distance between
every other layer cannot be smaller than 2, which is the nearest
distance between centers of two superballs aligned along one of
the coordinate directions. As $p$ decreases from unity, there
exists a point $p^*_o$ where the particles of every other layer
contact each other, forming ``cages'' for the particles of the
layer in between. The value of $p^*_o$ can be obtained from the
relation.

\begin{equation}
\label{eq6} 2\left ({1-2^{1-2p}}\right )^{\frac{1}{2p}} - 1=0,
\end{equation}

\noindent and we have $p^*_o = \ln3/\ln4 = 0.7924\ldots$ with the
corresponding packing density $\phi^*_o= 0.7959\ldots$. If $p$
decreases further (i.e., $p<p^*_o$), the ``cages'' will be too large
for the superballs to maintain 12 contacts and the packing density
would decrease.

Thus, the $\mathbb{O}_0$ lattice only gives the densest known packings
of superballs in the vicinity left of the sphere point, i.e.,
$p^*_o<p<1$. The packing density is given by

\begin{equation}
\label{eq7} \displaystyle{\phi_{max} = \frac{V_{sb}(p)}{|{\bf
e}_1\times{\bf e}_2\cdot{\bf e}_3|} = \frac{V_{sb}(p)}{8\left
({1-2^{1-2p}}\right )^{\frac{1}{2p}}}},
\end{equation}

\noindent where $V_{sb}(p)$ is the volume of a superball given by
Eq.~(\ref{eq4}). Packing density $\phi_{max}$ as a function of $p$
is plotted in Fig.~\ref{fig4} (the branch with $p^*_o<p<1$). The ``left''
slope of $\phi_{max}$ at $p=1$ is given by

\begin{equation}
a_- =  -\frac{\pi}{12\sqrt2} \left[{\ln\left({64}\right)+{8+
4\Psi\left(\frac{1}{2}\right) + 2\Psi \left(\frac{3}{2}\right) -
6\Psi \left(\frac{5}{2}\right)} }\right] = -0.02941\ldots.
\end{equation}

\noindent Thus, the initial increase of the packing density as $p$
decreases from unity is linear in $(1-p)$. As pointed out in the
previous section, the linear dependence on $|p-1|$ indicates the
nonanalyticity of $\phi_{max}$ at $p=1$, resulting from the broken
symmetry of superballs which is distinctly different from other
known hard particle systems such as ellipsoids in three
dimensions. Note that the magnitude of $a_-$ is smaller than
$a_+$, implying that superballs in the cubic regime are more
efficient at filling space than superballs in the octahedral
regime, as can be also seen in the limiting cases: cubes can
tile space while the densest known packing of regular octahedra
covers about $94.74\%$ of the space. The fact that the
density of the $\mathbb{O}_0$ lattice packing begins to decrease when
$p<p^*_o$ suggests the existence of packings with different symmetry
that would be optimal near the octahedron point.

\subsection{The $\mathbb{O}_1$-Lattice Packing}

\begin{figure}
\begin{center}
$\begin{array}{c@{\hspace{0.75cm}}c@{\hspace{0.75cm}}c}\\
\includegraphics[height=4.5cm, keepaspectratio]{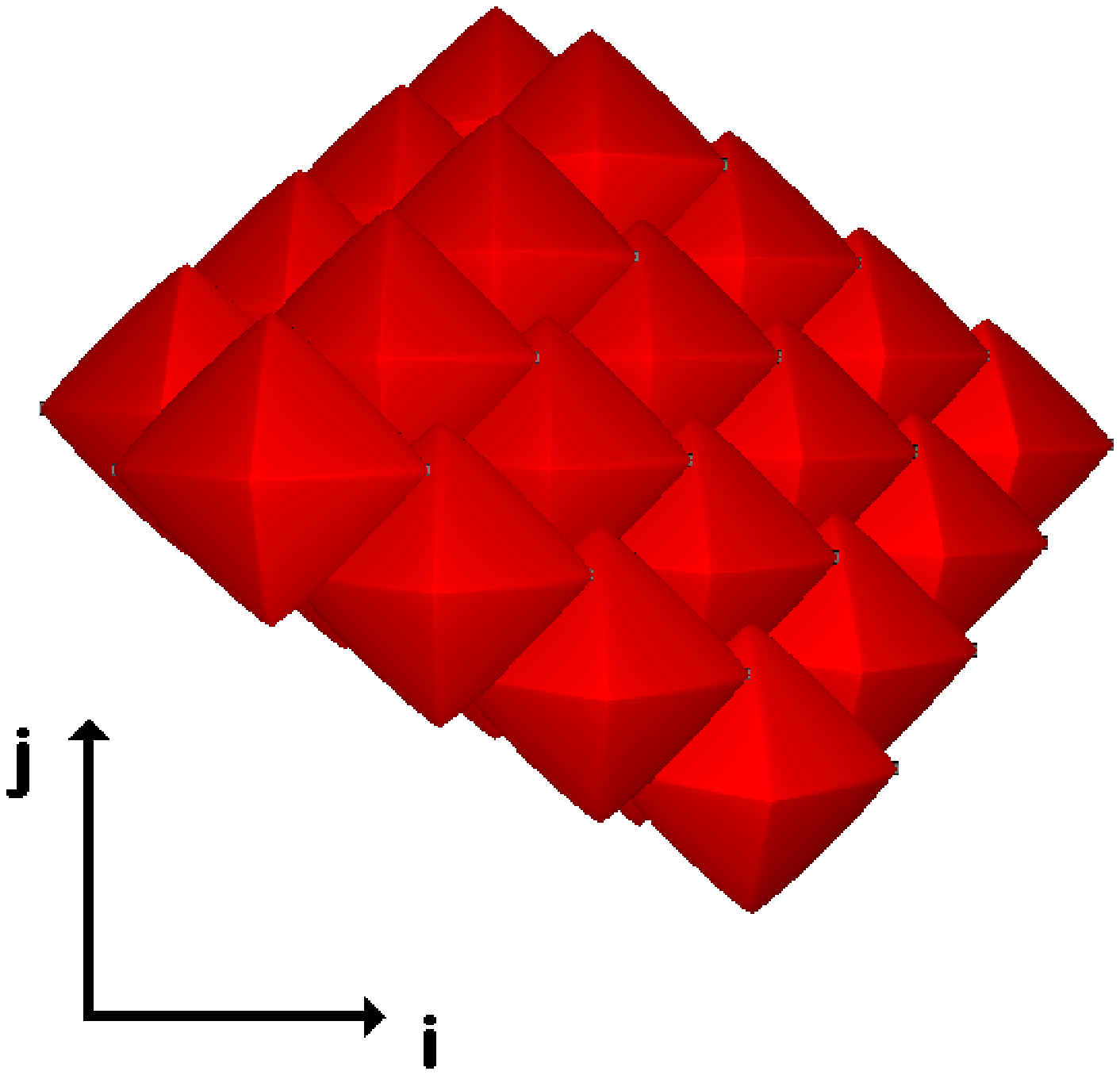}&
\includegraphics[height=4.5cm, keepaspectratio]{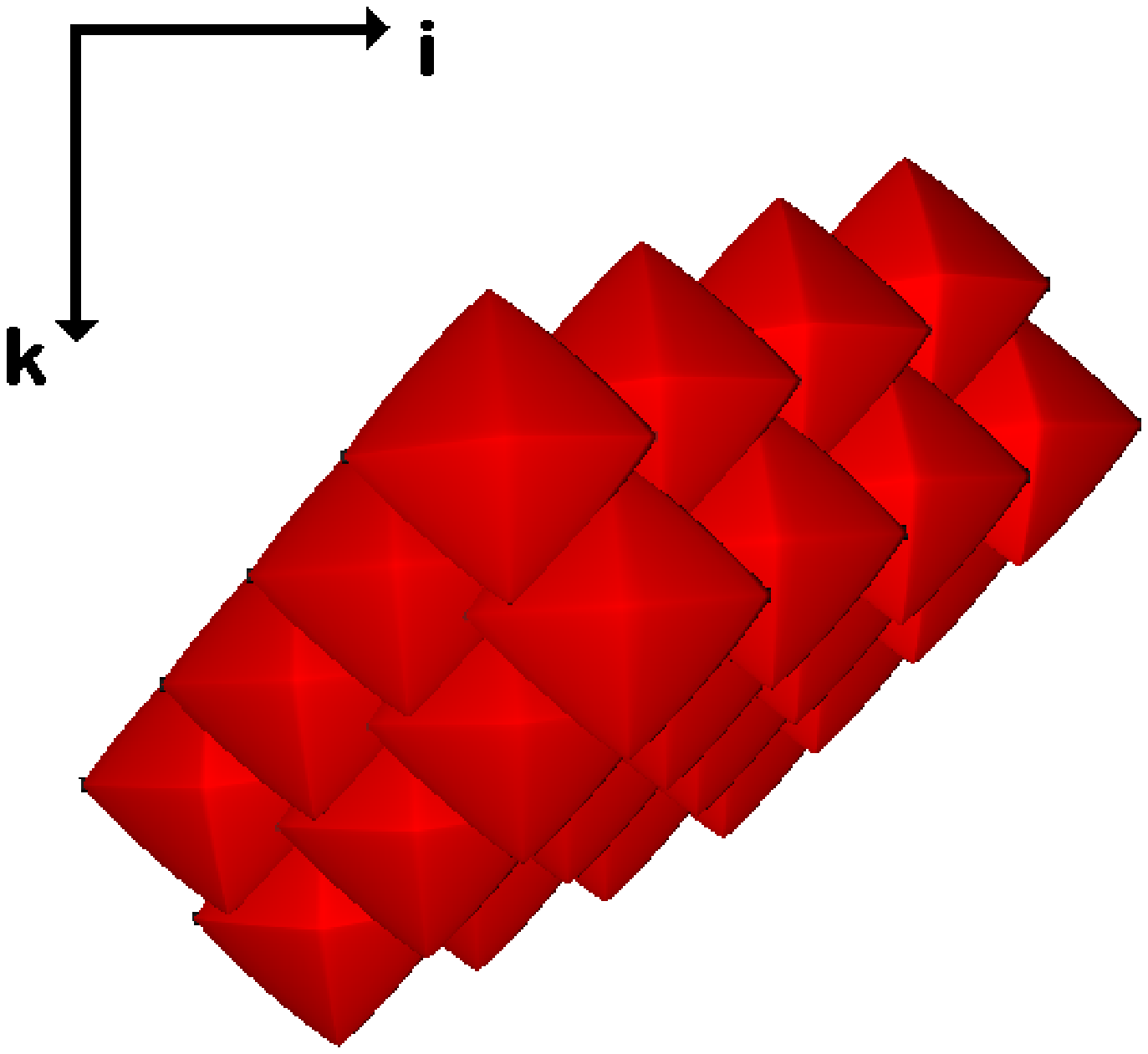}&
\includegraphics[height=4.5cm, keepaspectratio]{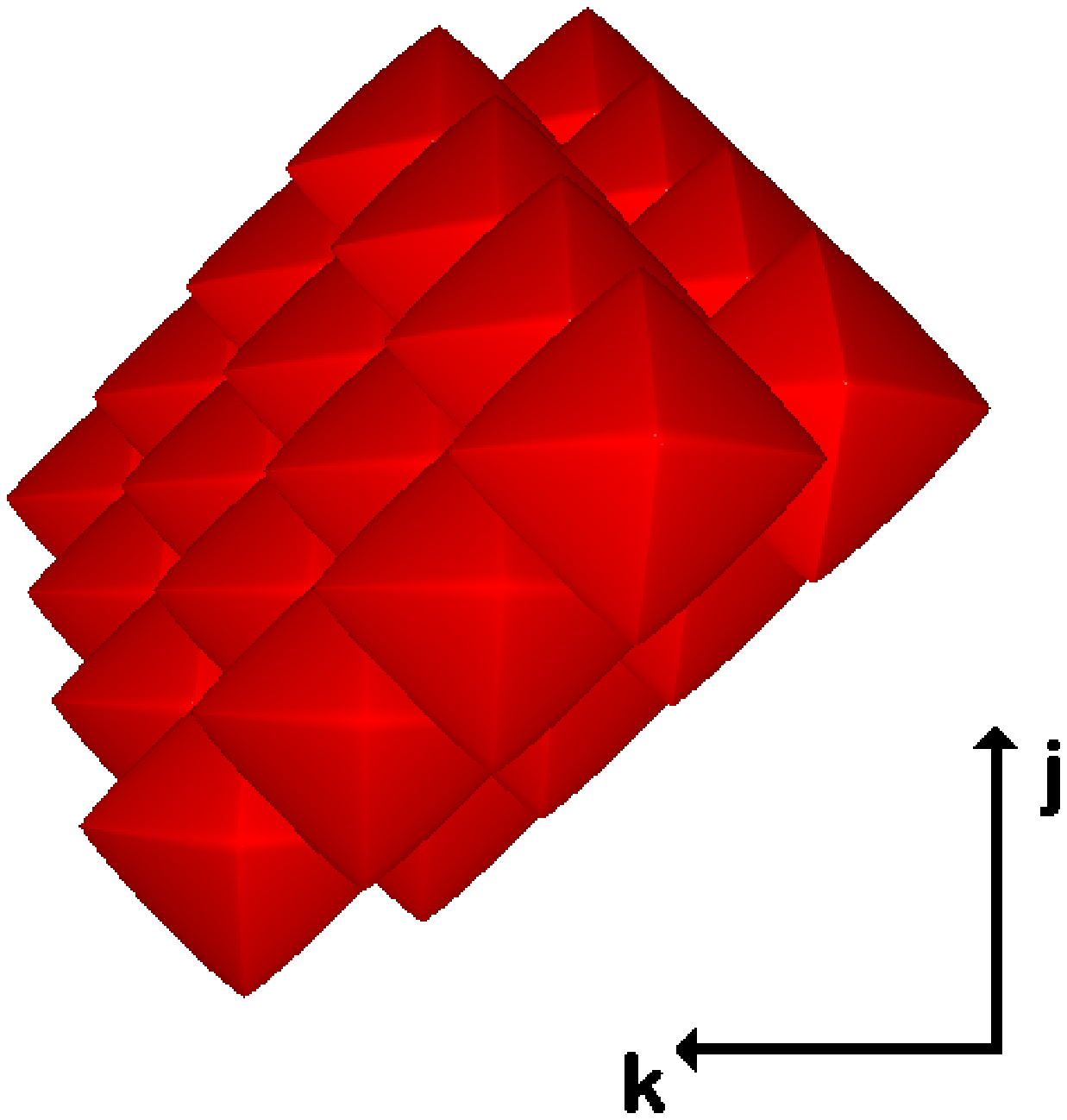}\\
\mbox{\bf (a)} & \mbox{\bf (b)} & \mbox{\bf (c)}
\end{array}$
\end{center}
\caption{(color online). Projections of a $\mathbb{O}_1$-lattice
sub-packing on three orthogonal coordinate planes: (a) Projection
on the (001) plane. (b) Projection on the (010) plane. (c)
Projection on the (100) plane.} \label{fig10}
\end{figure}

 In particular, the densest known
packing of regular octahedra, defined by $|x_1|+|x_2|+|x_3|\le1$, is
achieved by a Bravais lattice packing with the lattice vectors \cite{Henk}

\begin{equation}
\label{eq10} \displaystyle{\bf e}_1 = \frac{2}{3}{\bf
i}+\frac{2}{3}{\bf j}-\frac{2}{3}{\bf k}, \quad {\bf e}_2 =
-\frac{1}{3}{\bf i}+\frac{4}{3}{\bf j}-\frac{1}{3}{\bf k}, \quad
{\bf e}_3 = \frac{1}{3}{\bf i}-\frac{1}{3}{\bf j}-\frac{4}{3}{\bf
k}.
\end{equation}
This packing, with density $\phi_{M} = 18/19 = 0.9473\ldots$, 
was discovered by Minkowski \cite{Mink}.  By perturbing the Minkowski lattice packing using the
method depicted in Sec.~II.B, we are able to identify a family of
highly dense packings of superballs near the octahedron point,
i.e., the $\mathbb{O}_1$-lattice packing [see Fig.~\ref{fig5}(d)]. As shown
in Fig.~\ref{fig10}, the projections of the $\mathbb{O}_1$-lattice packing on
the three orthogonal coordinate planes possess certain
translational symmetry in two dimensions. The translational
vectors are along $({\bf i}+{\bf j})$ direction for (001) plane,
along $({\bf i}+{\bf k})$ and $({\bf k}-{\bf i})$ directions for
(010) plane and along $({\bf k}-{\bf j})$ direction for (100)
plane. The magnitude of the translational vectors are functions of
$p$, which can be determined with the additional condition that
each particle has 12 contacting neighbors. Thus, we obtain the
lattice vectors for the $\mathbb{O}_1$, i.e.,

\begin{equation}
\label{eq11} {\bf e}_1 = 2l{\bf i}+2l{\bf j}-2l{\bf k}, \quad {\bf
e}_2 = -2q{\bf i}+2s{\bf j}-2q{\bf k}, \quad {\bf e}_3 = 2q{\bf
i}-2q{\bf j}-2s{\bf k},
\end{equation}

\noindent where $l$, $s$ and $q$ are given by the following
equations

\begin{equation}
\label{eq12}
\begin{array}{c}
\displaystyle{l=3^{-\frac{1}{2p}}, \quad\quad 2q^{2p}+s^{2p}-1=0,} \\
\displaystyle{(l+q)^{2p}+(s-l)^{2p}+(l-q)^{2p}-1 = 0.}
\end{array}
\end{equation}

\noindent At the limit $p=0.5$, Eq.~(\ref{eq11}) reduces to
Eq.~(\ref{eq10}).

As $p$ increases from 0.5, the superballs deviate more and more
from regular octahedra and it becomes more difficult to maintain
12 contacts for each particle and simultaneously to keep the
translational symmetry of the projected packings. Interestingly,
as $p$ exceeds $p^*_o = \ln3/\ln4$, Eq.~(\ref{eq12}) will have no
real roots, indicating that the symmetry and contacting conditions
could not be satisfied at the same time beyond $p^*_o$. Thus,
the $\mathbb{O}_1$ lattice gives the densest known packings of
superballs near the octahedron point, i.e., $0.5<p<p^*_o$. At $p^*_o$,
the $\mathbb{O}_0$- and $\mathbb{O}_1$-lattice packings have exactly
the same density, i.e., $\phi^*_o= 0.7959\ldots$ and
superballs with $p=p^*_o$ possess a two-fold degenerate crystalline
maximal density state. Note that the $\mathbb{O}_0$ and $\mathbb{O}_1$
lattices possess distinct symmetries, as $p$ passes $p^*_o$ the
apparently optimal packing ``jumps'' from one lattice to the
other, which results in one more nontrivial nonanalytic point in
$\phi_{max}$.  This additional nonanalytical point does not exist
for superdisk packings. Nor does similar behavior exist for 
any other known hard-particle systems.

We emphasize that
the $\mathbb{O}_1$- and $\mathbb{O}_0$-lattice packings are
\textit{separately} constructed from the FCC packing of spheres
which has been proved to be optimal, and from the Minkowski
packing -- the densest known packing of regular octahedra,
respectively. The amazing fact that the two families of packings
both terminate at $p^*_o$ with the same density suggests that these
packings are very likely optimal, although we could not completely
exclude the possibility that a periodic packing with a complex
particle basis might have a higher density over an interval
including $p^*_o$. In other words, the existence of the continuous ``path'' of 
superball packings that we found connecting the FCC lattice packing of spheres and 
the Minkowski lattice packing of regular octahedra given the optimal packings 
of two limiting particle shapes strongly suggests that the packings along the path are also optimal.

\begin{figure}
\begin{center}
$\begin{array}{c@{\hspace{0.75cm}}c@{\hspace{0.75cm}}c}
\includegraphics[height=4.5cm, keepaspectratio]{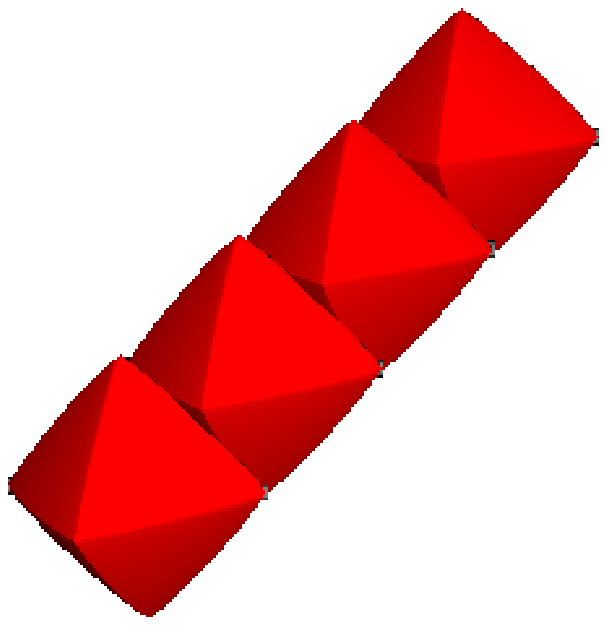} &
\includegraphics[height=4.5cm, keepaspectratio]{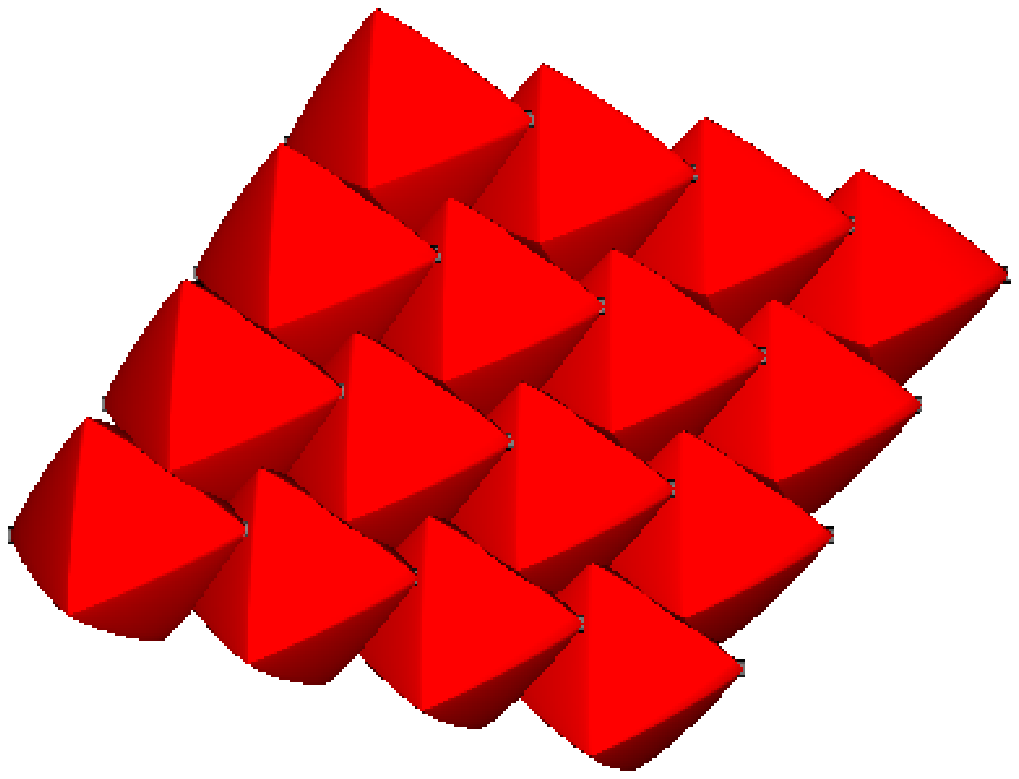} &
\includegraphics[height=4.5cm, keepaspectratio]{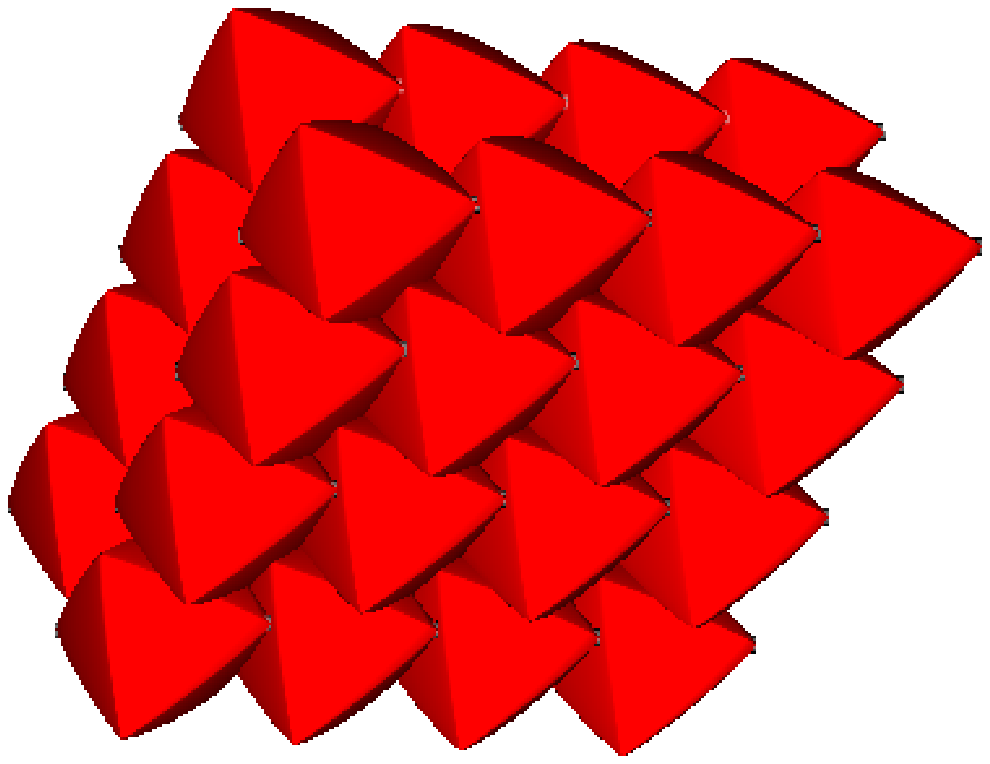} \\
\mbox{(a)} & \mbox{(b)} & \mbox{(c)}
\end{array}$
\end{center}
\caption{(color online). The $\mathbb{O}_1$-lattice packing of
superballs obtained by stacking superball chains: (a) The
superball chain. (b) The layer constructed by stacking the chains.
(c) The packing obtained by stacking the layers.} \label{fig11}
\end{figure}

The packing density $\phi_{max}$ is given by

\begin{equation}
\label{eq13} \displaystyle{\phi_{max} = \frac{V_{sb}(p)}{|{\bf
e}_1\times{\bf e}_2\cdot{\bf e}_3|} =
\frac{V_{sb}(p)}{8l(3q^2+s^2)}},
\end{equation}

\noindent where $V_{sb}(p)$ is the volume of a superball given by
Eq.~(\ref{eq4}). The plot of $\phi_{max}$ as a function of $p$ is
shown in Fig.~\ref{fig4} (the branch with $0.5<p<p^*_o$). The packing
structure could also be considered as the stacking of superball
chains, as illustrated in Fig.~\ref{fig11}.

\subsection{Generalization to Concave Superballs}

\begin{figure}
$\begin{array}{c}
\includegraphics[height = 6.5cm, keepaspectratio]{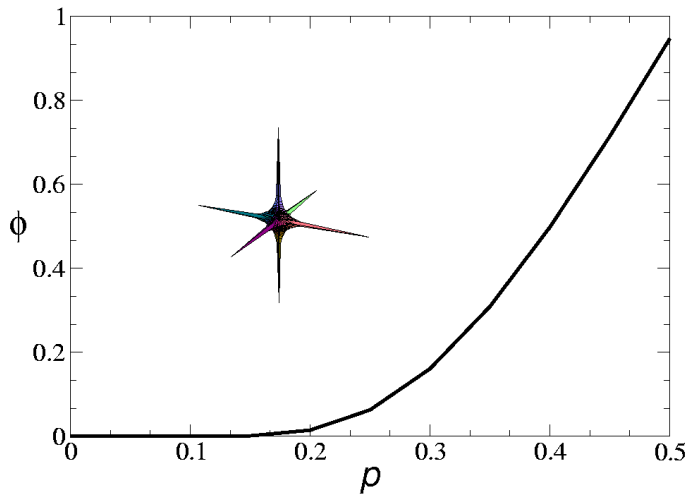}
\end{array}$
\caption{(color online). Density versus  deformation parameter $p$
for the packings of concave superballs. Insert: a concave
superball with $p = 0.1$, which will becomes a three-dimensional
cross at the limit $p \rightarrow 0$.} \label{fig12}
\end{figure}

As $p$ decreases from 0.5, the superballs become concave particles,
but they still possess octahedral-like shapes [see Fig.~\ref{fig2}(a)]. The lack of 
simulation techniques to generate concave superball packings
make it very difficult to find the optimal packings for the entire
range of concave shapes ($0<p<0.5$). However, based on our conclusions for
convex superball packings, we conjecture that near the octahedron
point, the optimal packings possess similar translational symmetry
in two dimensions to that of the $\mathbb{O}_1$-lattice packing. By
inserting into each octahedron in the Minkowski packing a concave
superball, we are able to obtain a family of packings that are
dense near $p=0.5$ but not optimal, the density of which is shown
in Fig.~\ref{fig12}. The packings could be densified a little by carefully
adjusting the particles to minimize the exclusion volume effects.
However, the adjustments are difficult to quantify.

It is interesting to point out that at the limit $p \rightarrow
0$, one can construct zero-length ``chains'' of a infinite number
of ``crosses'' along the [111] (or any equivalent) direction. In
other words, the centers of ``crosses'' packed along the chain are
infinitely close to each other. This aligning effect, which 
appears in three dimensions but not in two dimensions, suggests that near the ``cross''
point, the similarly constructed chains of superballs might be a
reasonable building block to produce dense packings.

\section{Conclusions}


In this paper, we have constructed the densest known packings of
convex superballs based on symmetry and contacting
requirements. For superballs in the cubic regime ($p>1$), the
candidate optimal packings are achieved by the $\mathbb{C}_0$- and 
$\mathbb{C}_1$-lattice
packings possessing two-fold and three-fold rotational symmetry, 
respectively, which can both be
considered as a continuous deformation of the FCC lattice. For
convex superballs in the octahedral regime ($0.5<p< 1$), the $\mathbb{O}_1$- and
$\mathbb{O}_0$-lattice, obtainable from continuous deformation of
the FCC lattice keeping its four-fold rotational symmetry, and
from the Minkowski lattice for regular octahedra \cite{Mink}
keeping the translational symmetry of the projected lattice on the
coordinate planes, are apparently optimal in the vicinity of
the sphere- and the octahedron-point, respectively. 
The packings jump between distinct lattices as $p$ passes $p^*_c$  and $p^*_o$, 
leading to two nontrivial nonanalytic points in $\phi_{max}$.

The existence of the continuous ``path'' of 
superball packings that we found connecting the FCC lattice packing of spheres and 
the Minkowski lattice packing of regular octahedra provides strong 
evidence that our candidate packings are very likely optimal.  
For concave superballs, we constructed a family of
dense packings based on the Minkowski lattice and discussed an
interesting aligning effect. Moreover, we have shown that as $p$
changes from unity, there is a significant increase of the
maximal packing density $\phi_{max}$, which is initially linear in
$|p-1|$ resulting in the nonanalytic behavior of $\phi_{max}$ at the
sphere point. Interestingly, this three-dimensional behavior
is  consistent with the fact that there is  an exponential improvement 
on the lower bound on $\phi_{max}$ of superballs relative to that for spheres in arbitrarily high
dimensions as found by Elkies, Odlyzko and Rush \cite{Elkies}. The packing characteristics of
superball packings are much richer than those of superdisks, and the
nontrivial influences of the broken symmetry are distinctly
different from the other known aspherical packings, such as 
ellipsoid packings. 
Although the anisotropic shape of the particle allows for the
possibility of local arrangements with larger coordination
numbers, in the candidate optimal packings each superball
maintains twelve contacting neighbors for all values of $p$, in
contrast to the densest known ellipsoid packings in which each
particle has fourteen contacting neighbors.

The optimal Bravais lattice packings of convex superballs for
a particular value of $p$ are also the corresponding dense
crystal phase states of superballs in equilibrium. Therefore, our
findings provide a starting point to quantify the entire equilibrium 
phase behavior of superball systems, which should deepen our
understanding of the statistical thermodynamics of non-spherical
particle systems \cite{Frenkel,Burton,Ken}. Ideally, one would expect a first-order
disorder-order phase transition of superball systems for every
value of $p$. However, the kinetic difficulty of such a transition
increases with the geometrical frustrations introduced by the 
particles. In practice, the disordered metastable states could frequently occur,
as illustrated in Fig.~\ref{fig3}(b). Some preliminary investigations 
that we have carried out show
that  disordered superball packings are {\it hypoconstrained},
i.e., the average contact number is smaller than twice of the
number of degrees of freedom of the particles \cite{AlexksIV}. A rotational
transition is also possible, i.e., the rotational degrees of
freedom of the particles are suddenly frozen when the packing
density exceeds a certain threshold value. A unique feature of 
superball systems is that \textit{symmetry transitions} occur at
$p^*_c$ and $p^*_o$. Thus for particles with a deformation parameter
close to $p^*_c$ or $p^*_o$, multi-morphological maximal density states might
be found.

Moreover, the optimal packings of convex superballs are
significantly denser than optimal sphere packings when $p$
deviates considerably from unity; even the disordered packings
could have a densities close to $\phi_{max}$ of spheres. This
suggests that one could use particles with shapes similar to that of
superballs in situations where dense packings are favored, such as
in certain powder sintering processes. We also note that 
superballs can be mass produced using modern lithography
techniques, which would enable one to prepare and study superball
packings experimentally.

In the future, we will generalize the current work to the study of
disordered packings of superdisks and superballs and further
explore the effect of broken rotational symmetry. It will be also interesting to
understand the ``jamming" characteristics \cite{SalJPC, AlexksIV}
of both random and crystalline packings of these particles.

\appendix

\section{Optimality of the Superdisk Packings Found in Ref. \cite{Superdisk}}

In this section, we briefly outline the procedure used to verify that 
for every specific value of $p>0.5$ the constructed packings of
superdisks in two dimensions in Ref.~\cite{Superdisk} are indeed
optimal, by virtue of Fejes Toth's theorem \cite{Fe64,Pach} described in Sec. I. In
particular, Fejes Toth's theorem states that the densest packing of any
centrally symmetric convex body in two dimensions is always given
by the smallest hexagon that circumscribes the body. Since the
body is centrally symmetric, the hexagon is as well and therefore
it tiles $\mathbb{R}^2$. Fejes Toth's theorem also implies that the
optimal packing for any two-dimensional centrally symmetric body
is a Bravais lattice packing. Thus, the problem of finding the
optimal packing reduces to finding the circumscribing hexagon of
a superdisk with minimum area.

\begin{figure}
\label{fig15}
$\begin{array}{c}
\includegraphics[height = 5.5cm, keepaspectratio]{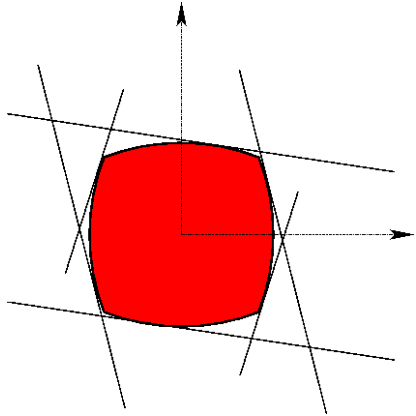}
\end{array}$
\caption{(color online). A superdisk circumscribed by a hexagon
formed from three pairs of parallel lines.}
\end{figure}

Let a superdisk be circumscribed by three pairs of parallel lines
forming a hexagon, as shown in Fig.~16. Denote the common tangent
points by $M_i$ with the coordinates $(m^{(i)}_1, m^{(i)}_2)$
($i=1,...,6$). Since the particle is centrally symmetric, $M_i$
and $M_{i+3}$ $(i=1,2,3)$ are also symmetric about the center,
i.e., $m^{(i)}_j = -m^{(i+3)}_j$ ($j=1,2$).

From the common tangent conditions, one can write down the
equations of the parallel lines in terms of the coordinates
$m^{(i)}_j$ ($i=1,...,6$, $j=1,2$), from which the vertices of the
hexagon as intersections of different pairs of lines can be
obtained. Thus, the area of the hexagon is readily expressed as a
function of $m^{(i)}_j$, i.e., $S(m^{(1)}_1, ..., m^{(6)}_2)$. By
solving the following optimization problem

\begin{equation}
\begin{array}{c}
Minimize: \quad S(m^{(1)}_1, ..., m^{(6)}_2), \\
Subject~to:
\quad|m^{(i)}_1|^{2p}+|m^{(i)}_2|^{2p}=1,~~(i=1,...,6);
\end{array}
\end{equation}

\noindent one can find the smallest hexagon circumscribing the
particle and thus verify for every specific $p>0.5$, the packings
given in Ref.~\cite{Superdisk} are indeed optimal.

\begin{acknowledgments}
The authors would like to thank Aleksandar Donev, Martin Henk and Greg Kuperberg for valuable
discussions. S. T. thanks the Institute for Advanced Study for
its hospitality during his stay there.
This work was supported by the Division of Mathematical Sciences
at the National Science Foundation under Award Number DMS-0804431
and by the MRSEC Program of the
National Science Foundation under Award Number DMR-0820341.
\end{acknowledgments}


\begin{thebibliography}{31}
\bibitem{supra}
Supramolecular chemistry deals with the chemistry and collective 
behavior of molecular building blocks that are organized on
large length scales (relative to molecular sizes)
with long-range order.
\bibitem{prelim} It is highly likely that the largest density 
of tetrahedral packings reported in
Ref. \cite{BChen} can be improved.
\bibitem{shape}
There is a fundamental distinction between the shapes of superball surfaces in the 
octahedral-like regime ($0 \le p < 1$) and the cubic-like regime ($1 < p \le +\infty$).  
The former display a set of linear loci along which a principal curvature diverges, representing a generalization of the edge set for the true octahedron at $p=1/2$.  No such loci of diverging curvature exist for the cubic-like regime, except in the  limit $p \rightarrow +\infty$.
\bibitem {footnote1}
Two polyhedra are dual to each other if the vertices 
of one correspond to the faces of the other.
\bibitem{footnote2}
The regular octahedron is centrally symmetric. In the densest octahedral lattice packing
found by Minkowski \cite{Mink}, 
the octahedra have the maximal number of face-to-face contacting neighbors, which 
significantly reduces the free volume. This observation together with 
the results of our optimization simulations described in the text strongly suggests 
that the Minkowski lattice packing of regular octahedra is optimal
among all packings of octahedra, which we conjecture to be the case.
\bibitem{sat}
In a saturated packing, there is no space available to add another
particle to the packing.
\end{thebibliography}


\end{document}